\documentclass[graybox, secnum]{svmult}

\usepackage{mathptmx}       
\usepackage{helvet}         
\usepackage{courier}        
\usepackage{type1cm}        
\usepackage{makeidx}         
\usepackage{graphicx}        
\usepackage{multicol}        
\usepackage[bottom]{footmisc}
\usepackage{hyperref}        
\usepackage{soul}            
\hypersetup{colorlinks=true,urlcolor=blue}
\usepackage[square,numbers]{natbib}
\makeindex             
\usepackage{amssymb}

\newcommand{\lsim }{{\lower0.8ex\hbox{$\buildrel <\over\sim$}}}
\newcommand{\gsim }{{\lower0.8ex\hbox{$\buildrel >\over\sim$}}}
\newcommand{\Msun}{\ifmmode {M_{\odot}}\else${M_{\odot}}$\fi}
\newcommand{\Lsun}{\ifmmode {L_{\odot}}\else${L_{\odot}}$\fi}
\newcommand{\Rsun}{\ifmmode {R_{\odot}}\else${R_{\odot}}$\fi}
\newcommand{\ergs}{erg~s$^{-1}$}
\newcommand{\ergcms}{erg~s$^{-1}$~cm$^{-2}$}

\newcommand{\lx}{L_{\mathrm{X}}}
\newcommand{\ledd}{L_{\mathrm{Edd}}}
\newcommand{\dist}{(D/8~\mathrm{kpc})^2}

\newcommand{\bepposax}{\textit{BeppoSAX}}
\newcommand{\swift}{\textit{Swift}}
\newcommand{\xrt}{\textit{Swift}/XRT}

\newcommand{\chandra}{\textit{Chandra}}

\newcommand{\xmm}{\textit{XMM-Newton}}
\newcommand{\integral}{\textit{INTEGRAL}}
\newcommand{\rosat}{\textit{ROSAT}}
\newcommand{\maxi}{\textit{MAXI}}
\newcommand{\rxte}{\textit{RXTE}}

\begin{document}
\title*{Low-Mass X-ray Binaries}
\author{Arash Bahramian \thanks{corresponding author} and Nathalie Degenaar}
\institute{Arash Bahramian \at International Centre for Radio Astronomy Research – Curtin University, GPO Box U1987, Perth, WA 6845, Australia, \email{arash.bahramian@curtin.edu.au}
\and Nathalie Degenaar \at Anton Pannekoek Institute for Astronomy, University of Amsterdam, Postbus 94249, NL-1090 GE Amsterdam, the Netherlands, \email{degenaar@uva.nl}}
%
%
\maketitle


\abstract{
A large fraction of X-ray sources in our Galaxy are low-mass X-ray binaries, containing a black hole or a neutron star accreting from a gravitationally bound low-mass ($\lsim$1 M$_\odot$) companion star. These systems are among the older population of stars and accreting systems in the Galaxy, and typically have long accretion histories. Low-mass X-ray binaries are categorized into various sub-classes based on their observed properties such as X-ray variability and brightness, nature of the companion star and/or the compact object, and binary configuration. In this Chapter, we review the phenomenology of sub-classes of these systems and summarize  observational finding regarding their characteristics, populations, and their distribution in the Galaxy.}
\section*{Keywords} 
Accretion Discs; Black Holes; Neutron Stars; X-ray Binaries.

\tableofcontents

\section{Introduction}
In this Chapter we introduce and review basic properties of the low-mass X-ray binaries (LMXBs): binary star systems comprised of a compact object, either a black hole (BH) or neutron star (NS), that accretes gas from a companion star less massive than the compact primary (typically $\lesssim$~1~M$_\odot$).\footnote{X-ray binaries with massive companion stars (typically $\gtrsim$~10~M$_\odot$) are called high-mass X-ray binaries (HMXBs).} The focus of this Chapter is on the demographics of LMXBs (e.g., binary configurations and formation, companion and compact object types, Galactic numbers and distribution), as well as their phenomenological behavior, particularly in the X-ray band (e.g., luminosity and variability).\footnote{Other Chapters in this book delve into the physics of accretion and compact objects. See Section~\ref{sec:xref} for a list of these chapters.}

\begin{figure}
    \centering
    \includegraphics[scale=0.1]{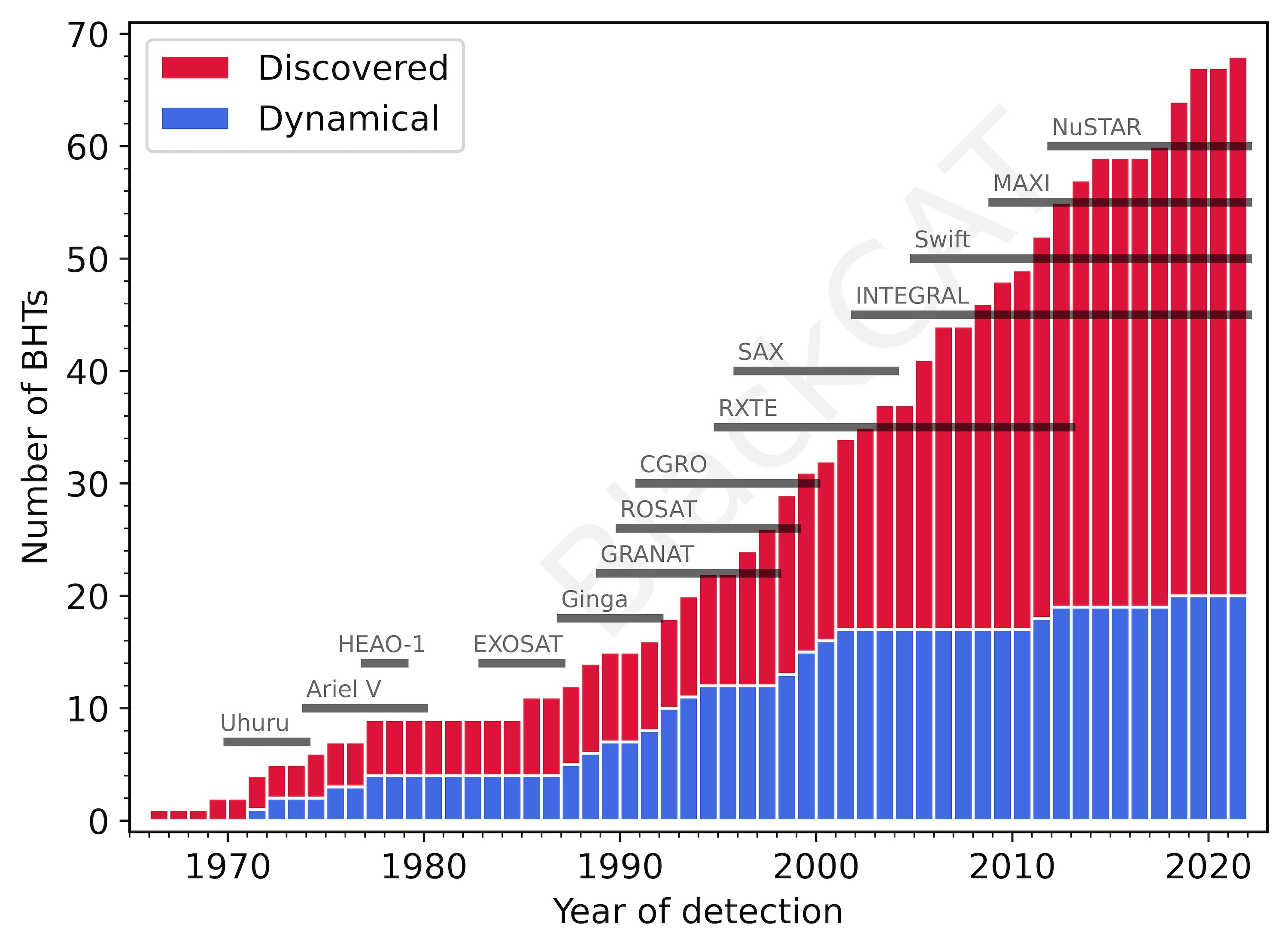}
    \caption{Growing population of BH LMXBs over the lifetime of X-ray observatories. Enhancements in observatories, and analysis has increased the rate at which transient X-ray binaries have been discovered/confirmed. This figure was created in December 2021 by the BlackCAT team, based on the continuously updated and publicly available data from BlackCAT \cite[][\url{https://www.astro.puc.cl/BlackCAT}]{CorralSantana16}.}
    \label{fig:blackcat}
\end{figure}

LMXBs are among the brightest X-ray point sources in the sky when they are actively accreting. This is therefore how they are typically discovered and identified. However, most LMXBs are not continuously accreting and are classified as transients, showing X-ray outbursts with high levels of accretion onto the compact object that typically only last for weeks to months. These accretion outbursts are separated by long intervals of quiescence, lasting months to tens of years, during which little or no accretion occurs. This transient behavior is commonly explained in terms of thermal-viscous instabilities in the accretion disk, the so-called Disk Instability Model (DIM; \cite{Smak1983,Lasota2000,hameury2020}).\footnote{See also chapter ``Formation and Evolution of Accreting Compact Objects'' by Belloni et al.} Despite their generally low duty cycles, enhancements in the sensitivity and monitoring capabilities of X-ray instruments have led to a steady increase in  the discovery of transient LMXBs (see Figure~\ref{fig:blackcat}).

To date, there are $\sim$200 confirmed and candidate LMXBs identified in our Galaxy, through their X-ray properties \citep[e.g.,][]{Liu2007, Tetarenko2016a, CorralSantana16}. For many LMXBs the nature of the compact primary remains to be identified, yet for several tens of systems we know whether these contain BHs or NSs. Observationally, known NS-LMXB outnumber BH-LMXBs by a factor of $\sim$2, and this is consistent with theoretical predictions based on formation history of NSs and BHs \citep{Kalogera98}. It is also worth noting that accreting BHs have been found more frequently in LMXBs, as opposed to HMXBs. This could either be a result of binary evolution \cite{belczynski2009}, or of observational biases \cite{casares2014_BeXRB}.

Around one fifth of the current observed NS-LMXBs sample is made up of accreting millisecond X-ray pulsars (AMXPs) and transitional Millisecond Radio Pulsars (tMSRPs). In these systems, the accretion flow near the NS is shaped by the magnetic field of the NS and is directed towards the magnetic poles, producing X-ray pulsations \cite{Bildsten1997,Wijnands1998}. AMXPs and tMSRPs are considered evolutionary links between (some) NS-LMXBs and recycled Millisecond Radio Pulsars (MSRPs).

In a large fraction of LMXBs, the compact object is accreting from a main sequence, sub-giant or red giant star that is filling its Roche lobe, and thus mass transfer towards the compact object is primarily through Roche lobe overflow. However, other types of LMXBs, for example, consisting of a compact object accreting from a stripped remnant ``core'' of a star, or winds of a low-mass giant that is not filling its Roche lobe (i.e., wind accretion) have also been observed. 

In this Chapter we briefly go into common methods to determine the nature of the compact object in LMXBs (Section~\ref{sec:NSBH}), after which we explore the dichotomy of LMXBs based on their companion type and accretion mechanism (Section~\ref{sec:donor}), X-ray luminosity and (long-term) X-ray variability (Section~\ref{sec:outbursts}), and their distribution throughout our Galaxy (Section~\ref{sec:distr_demo}). We close the Chapter with a brief summary and prospects offered by future observatories and surveys (Section~\ref{sec:summary}).

\section{The nature of the compact primary in LMXBs}\label{sec:NSBH}
Determining the nature of the compact object in a LMXB is generally an observationally challenging task. While there are phenomenological features in LMXB outbursts (either spectral or temporal) that sometimes allow distinguishing between NS- or BH-LMXB candidates \citep[e.g.,][]{Belloni05,Lin2007}, these features are not sufficient to prove the nature of the compact object. Here we review some of the more widely used methods to determine the nature of the compact object in LMXBs.

{\bf Bursts and Pulsations} - The most robust method for confirming the presence of a NS in a LMXB is the detection of events that require a solid surface, like thermonuclear X-ray bursts \citep[also called type-I X-ray bursts and arising from rapid runway fusion of accreted matter on the surface of the NS;][]{Lewin1993,Strohmayer2006}, or coherent (X-ray) pulsations \citep[caused by surface hotspots and modulated by the rotation of the NS;][]{Alpar1982,Bildsten1997}. NSs can be sub-classified based on their magnetic field. In LMXBs we find primarily weakly magnetic NSs with $B \lesssim 10^8$~G, though a handful of systems contain NSs with stronger magnetic fields of $B \lesssim 10^{11-13}$~G \cite{Patruno2012a}. However, NSs with strong magnetic fields are more commonly found in HMXBs \cite{Bildsten1997}. It is important to emphasize that the absence of coherent pulsations or thermonuclear X-ray bursts does not rule out a NS accretor in a LMXB, but there are other methods to investigate the nature of the compact primary. 

{\bf Radial velocity and estimation of the mass function} - One of the most robust methods for verifying BHs in LMXBs (and estimating their mass) is generally through radial velocity studies via optical or near-infrared spectroscopy - relying on Doppler displacement of spectral features that are confidently associated with the companion. While such radial velocity studies (and inferred mass estimates) have a strong dependence on systems' inclination angle - a parameter difficult to constrain observationally, a lower limit of $\geq 3$~M$_\odot$, unambiguously identifies the compact object as a BH, since degeneracy pressure would not be able to prevent a star of such mass from collapsing \citep[][]{vanParadijs1995,Casares2014}.

{\bf Disk-Jet coupling} - Another diagnostic method assisting in determining the compact object in LMXBs is based on the observed coupling between jet outflow (observed in the radio bands) and accretion rate (associated with the X-ray luminosity). While both NS- and BH-LMXBs show evidence for jets, when at similar X-ray luminosity, BHs tend to be brighter in the radio than NSs by a factor of $\sim$5 to 20 \citep{Fender2003, Migliari2006, Tudor2017, Gallo2018}. This correlation has shown to be promising in identifying candidates, but the scatter and overlap of NS and BH systems in this correlation is significant and thus it is not sufficient for confirming nature of the compact object. Two notable examples of this scatter are IGR~J17591$-$2342, a NS-LMXB that due to its bright radio emission was initially thought to harbor a BH \citep{Russell2018, Gusinskaia2020}, and the dynamically confirmed BH-LMXB Swift J1357.2$-$0933, for which its low radio luminosity initially led to speculations on the NS nature for the compact object \citep{Sivakoff2011ATel, MataSanchez2015}.

{\bf Quiescent X-ray properties} - The fact that NSs have a solid surface that, for typical temperatures of $\sim 10^{6}$~K, should give rise to a black-body like thermal emission component in the quiescent X-ray spectrum of the LMXB. This diagnostic has often been used, for example, to identify NS-LMXBs in globular clusters (see Section~\ref{subsec:globclust}). However, if the NS is very cold (causing its thermal emission peak to move out of the X-ray band), if residual accretion or magnetospheric processes play a role (giving rise to a bright, harder X-ray emission component), or if the extinction along the line of sight is very high (causing the soft thermal X-rays to be absorbed), this method may break down \citep[][]{Wijnands2005,Heinke2009,Degenaar2012a}. In other words, the absence of a thermal emission component in quiescent LMXBs is not evidence for a BH accretor.

{\bf Other methods} - Apart from the detection of surface phenomena, dynamical mass measurements, radio/X-ray luminosity ratio, or quiescent properties, other methods used to gauge the nature of the compact object in LMXBs include the ratio of their optical/infrared over X-ray fluxes \cite{Russell2006,Russell2007}, approximation of mass ratio via properties of the hydrogen H$_\alpha$ emission line \citep{Casares2016}, or their X-ray spectral-timing behavior \cite{wijnands1999,vanderKlis2006,Wijnands2015}. 

In exploring the nature of a compact object in a LMXB, it is important to note that detection of bursts or pulsations, and radial velocity studies generally lead to relatively robust classification of the compact object, while methods relying on luminosity ratios, emission features, and quiescent properties generally provide valuable circumstantial evidence regarding the nature of the compact object, particularly when the former methods are unfeasible.



\section{Donors and accretion phenomenology in LMXBs}\label{sec:donor}

Many physical and observational characteristics of LMXBs such as mass transfer rate, accretion mechanism, luminosity and outburst duty cycle are largely - but not exclusively, linked with the binary configuration properties like binary separation/period, eccentricity, and nature of the donor star, and, to a lesser extent, the nature of the compact object. For instance, all these factors play a role in the X-ray emission produced by a LMXB. The {\it intrinsic} brightness of a LMXB scales with the rate at which mass is accreted. There is a physical limit for the rate at which mass can be accreted hence on the brightness of LMXBs. This limit is the Eddington luminosity (above which radiation pressure prevents further accretion), which scales with the mass of the compact object as $L_{\mathrm{Edd}} \approx 10^{38}\left(M/M_{\odot}\right)$~\ergs. It is often useful to express the brightness of LMXBs as a fraction of this Eddington luminosity. 

\begin{figure}
    \centering
    \includegraphics[scale=0.205]{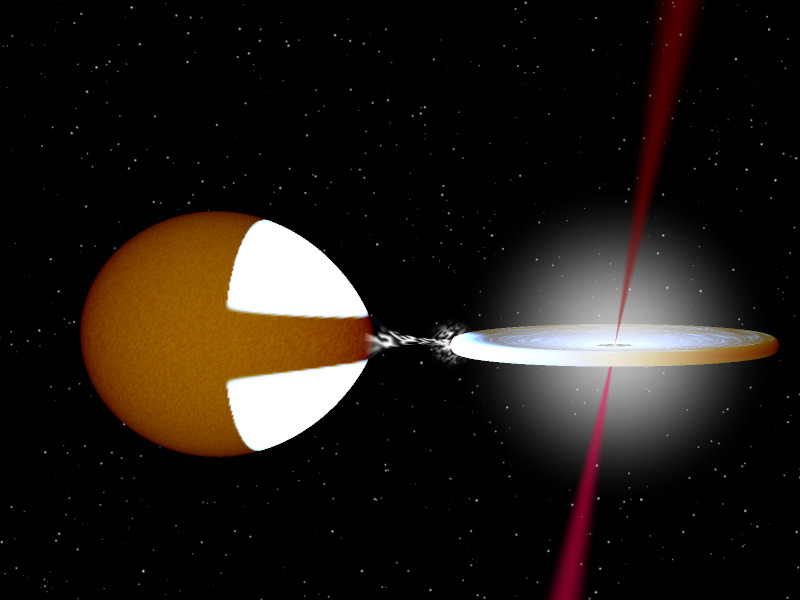}
    \includegraphics[scale=0.205]{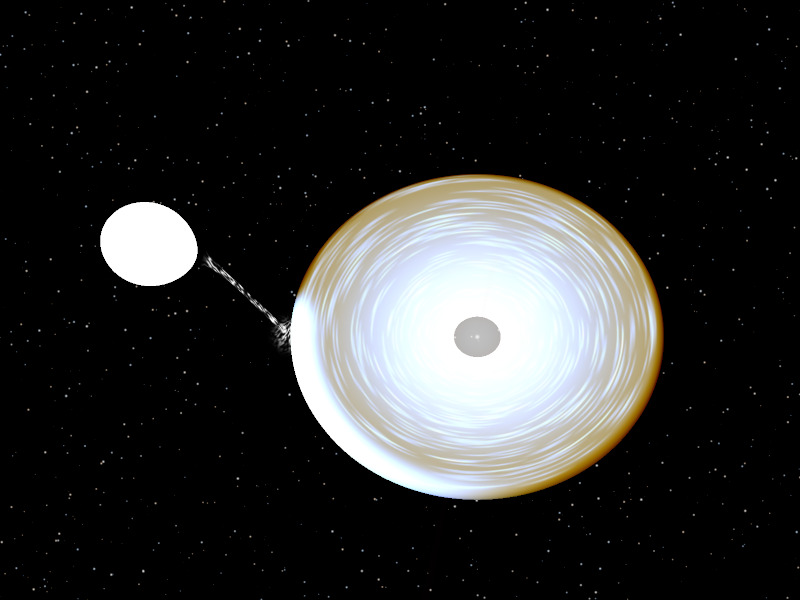}\\
    \includegraphics[scale=0.205]{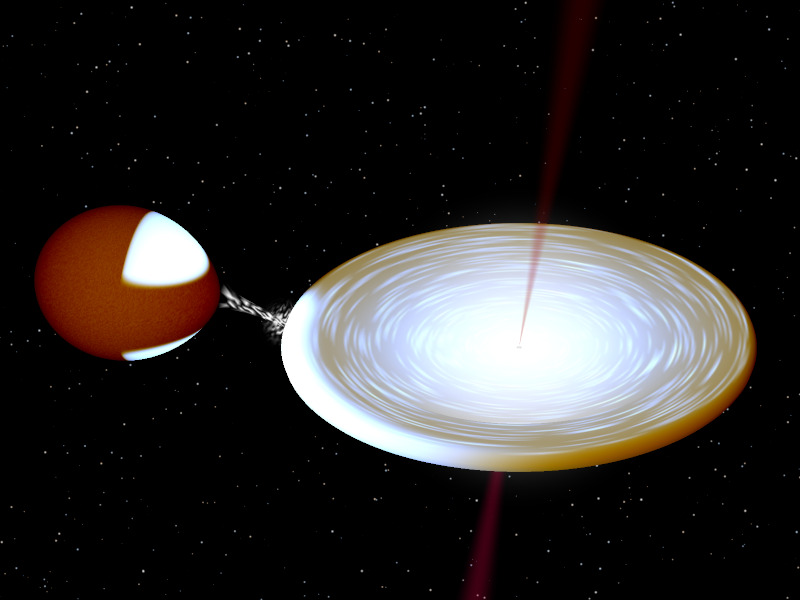}
    \includegraphics[scale=0.205]{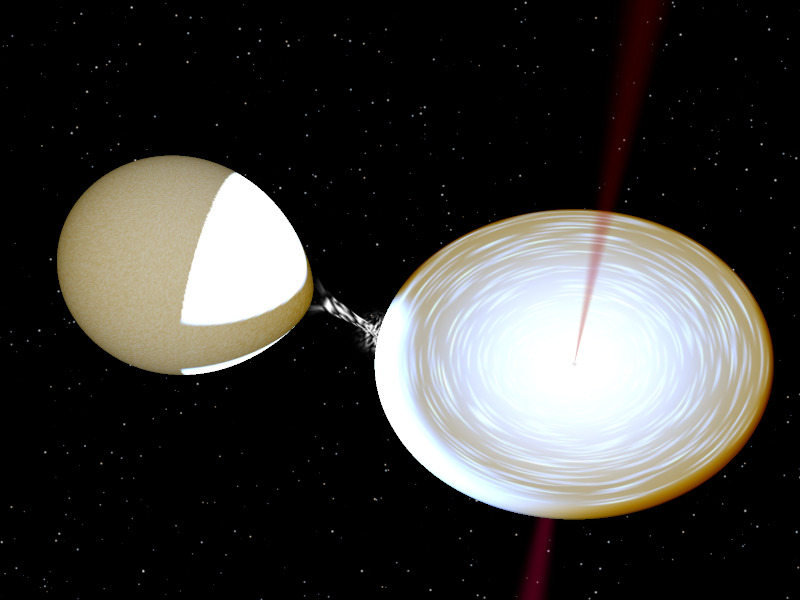}\\
    \includegraphics[scale=0.205]{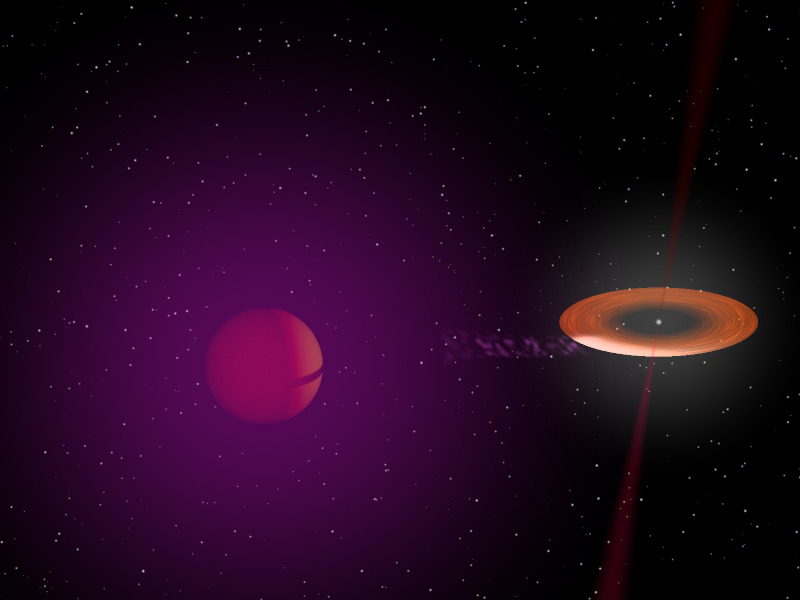}
    \includegraphics[scale=0.205]{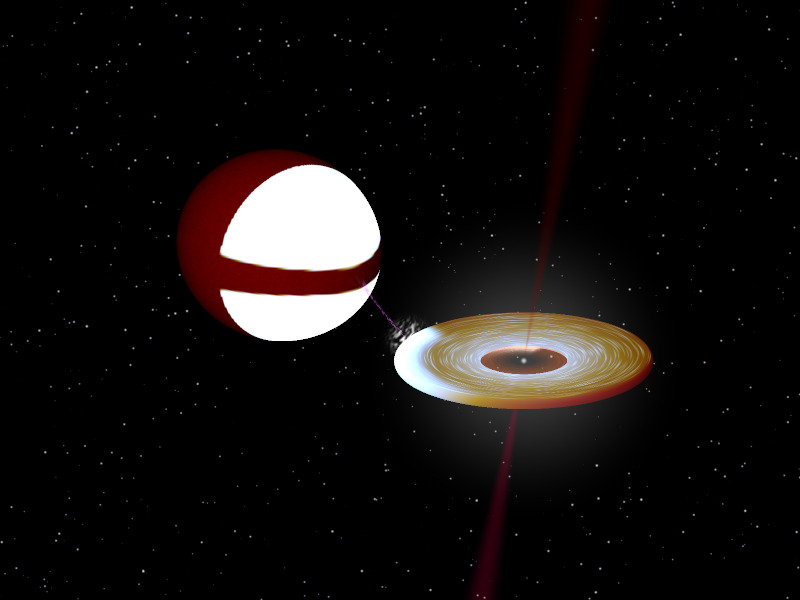}\\
    \caption{Illustration of LMXBs in various configurations. Top-left: a transient eclipsing LMXB during outburst (e.g., GRS~1747$-$312), top-right: a persistent ultra-compact X-ray binary with a low inclination angle (e.g., 2S~0918$-$549,\cite{Zhong2011}), middle-left: V404~Cyg during an outburst (as a well known transient BH-LMXB), middle-right: Scorpius~X-1 (as a well known persistent NS-LMXB), bottom-left: GC~1+4 (as a symbiotic X-ray binary), bottom-right: IGR~J17062--6143 (as a very faint X-ray binary). These visualisations were produced using the publicly available \textsc{binsim} code \cite[][\url{http://www.phys.lsu.edu/~rih/binsim/}]{Hynes2002}. Input for \textsc{binsim} includes parameters describing binary configuration (e.g., separation, inclination), companion properties (e.g., mass, Roche-lobe overflow fraction, temperature), accretion flow and disk characteristic (disk inner and outer radii, temperature gradient), jet angle and brightness, among other parameters. These visualizations prioritize simplification for clarity of impression over scientific accuracy. our aim here is to demonstrate relative size and brightness of some of the main components of LMXBs. Note that scales are not the same across the panels and vary from $\sim10^{-2}$~AU (e.g., for an ultra-compact X-ray binary) to $\sim$ a few AU (e.g., for a Symbiotic X-ray binary).}
    \label{fig:binsim}
\end{figure}

In this section, we discuss various sub-classes of LMXBs based on binary configuration as determined through observations, and their accretion properties. 
Figure~\ref{fig:binsim} shows graphical impressions of some of the different types of LMXB systems discussed in the next sections.

\subsection{Canonical Roche-lobe overflow with main sequence or giant stars}\label{subsec:RLOF}
A large fraction of LMXBs consist of main sequence, sub-giant or red giant branch stars filling their Roche lobe in a binary with a BH or a NS. Mass transfer from the companion then occurs through Roche-lobe overflow onto an accretion disk. The orbital period in these systems is typically of order of hours for systems with a main sequence star and order of days/weeks for systems with giants \citep{Liu2007, CorralSantana16}. Figure~\ref{fig:binsim} shows schematic impressions of different Roche-lobe overflow LMXBs (all but the bottom left image).

Most of these {\it canonical} LMXBs are transient systems, identified in the X-rays by their outbursts that typically reach $0.01 - 0.5  L_{\mathrm{Edd}}$ while in quiescence their X-ray luminosity is $\leq10^{-5} L_{\mathrm{Edd}}$. When actively accreting, the emission of LMXBs at all wavelengths is typically dominated by that of the accretion flow and associated outflows. Direct and detailed studies of the donor star are therefore generally not possible when active accretion is occurring (i.e., in outbursting transients or persistent LMXBs).

In quiescence, optical and infrared emission from LMXBs harboring main sequence stars and hydrogen-rich giants is generally dominated by the companion star, showing stellar atmospheric features like TiO features in the optical or CO bandheads in the near-infrared. These features are crucial for radial velocity measurements and thus dynamical mass measurements for compact objects in these systems. In addition to stellar features many of these systems show strong hydrogen emission lines (hydrogen H-$\alpha$ in the optical, or Bracket-$\gamma$ in the near-infrared), associated with the accretion disk. As these systems enter outburst, typically the optical and infrared emission gets dominated by the accretion disk continuum emission, sometimes with numerous hydrogen and helium emission features. The profile and strength of accretion emission lines H-$\alpha$ or Bracket-$\gamma$ (in quiescent and/or outburst) are among key indirect indicators for accretion geometry \citep{Steeghs2004} and potential nature of the compact object \citep{Casares2016}. 

\subsection{Ultra-compact X-ray binaries}\label{subsec:UCXB}
Ultra-compact X-ray binaries (UCXBs) are primarily defined (and identified) by their very short orbital period of $\lsim 80$~minutes. At these orbital periods, only helium burning sub-dwarf stars or white dwarfs are small enough to fit into the orbit and fill their Roche lobe \citep[][]{Paczynski81,Rappaport1982, Savonije1986}. We tabulate a list of currently known UCXBs and candidates in Table~\ref{tab:UCXBs}. A schematic impression of an UCXB is shown in Figure~\ref{fig:binsim} (top right and bottom right). 

Observationally, confirming the UCXB nature of a LMXB binary requires measurement of the orbital period. Thus, so far, only a few dozen UCXBs have been confirmed \citep[e.g.,][]{Cartwright2013}. While direct observation of orbital modulation has been the prime method for identifying UCXBs \citep[e.g.,][]{Stella1987, Zurek2009, Galloway2010, Zhong2011}, in a significant fraction of these systems, discovery of an accreting millisecond X-ray pulsar (AMXP) has allowed estimation of orbital period through accurate timing of the X-ray pulsations \citep[e.g.,][]{Wijnands1998,Chakrabarty1998_sax,Altamirano2010, Sanna2016, Strohmayer2018}. In addition to confirmed UCXBs, there are also candidate systems that are identified as such based on the X-ray evolution of their accretion outbursts \citep{Heinke2015,Stoop2021}, or their optical and X-ray spectral properties \citep{Bassa2006,ArmasPadilla2020, CotiZelati2021}. Furthermore, LMXBs that persistently accrete at very low rates are suspected to have small disks (since these are easier kept photo-ionized sustaining active accretion) and are hence considered good candidate UCXBs \cite{intZand2005a,intZand2007}.

While most UCXBs discovered to date have been found to be persistently bright in the X-rays with the X-ray luminosity varying typically at $L_X > 10^{36}$~\ergs, a subset of UCXBs are transient systems, showing quiescent periods with $L_X < 10^{32}$~\ergs\ interspersed by outbursts. Additionally, some UCXBs are classified as very faint X-ray binaries (see Section~\ref{subsubsec:vfxb}), showing very faint (transient) accretion emission of $L_X \simeq 10^{34}-10^{36}$~\ergs \citep{vanHaaften2012, Cartwright2013}. 

Given the natures of both accretor and companion in UCXBs, these systems are extremely hydrogen deficient and abundance of elements is particularly unusual, typically dominated by He, C, O, Ne (or less likely, by heavier elements). Thus the spectral continuum (particularly in the optical band) in these systems is generally void of stellar absorption features or any hydrogen emission or absorption lines. Instead, emission lines from He, N, C, O are prevalent in the optical \citep{Nelemans2004, Nelemans2006}, ultra-violet \citep{Homer2002} and X-rays \citep{Juett2001, Juett2003, Madej2010}.

To date, there have been no BHs confirmed in any UCXB system. However candidates have been identified based on X-ray and radio properties \citep{Bahramian2016}, X-ray outburst properties along with radio and X-ray properties \citep{Stoop2021}, or X-ray luminosity and optical spectroscopy \citep[in an extragalactic system,][]{Dage2019}. There might be an observational bias against detecting BH-UCXB systems in the X-rays because the radiative efficiency and duration of the outbursts of BH LMXBs may drop sharply towards shorter orbital periods \cite{Knevitt2014}. Furthermore, confirming the presence of a BH in a UCXB system dynamically is particularly challenging, as traditional methods, like radial velocity measurements based on spectral features associated with the surface of the donor, are unfeasible. 

Galactic compact binaries like UCXBs are expected to be among the main sources of gravitational wave emission to be detected by future facilities like the Laser Interferometer Space Antenna (LISA) and TianQin \citep{AmaroSeoane2017, Luo2016}. This includes the ``dual-line'' signal from the combination of NS spin and system orbit, and the ``chirp'' signal from those of these systems that eventually go through coalescence \citep{Nelemans2001, Tauris18, chen2021, Suvorov21}. Therefore, it is highly desired to search for and find more UCXBs.

\begin{table}
\centering
\begin{tabular}{lllccc}
\hline
\hline
System               & Distance& $P_\textrm{orb}$  & T/P   & NS/BH & Reference \\  
                     & (kpc)   &  (min)            &       &       &           \\
\hline
\multicolumn{6}{l}{{\bf Confirmed UCXBs}}\\
4U 1820$-$303        & 8.0     & 11       & P & NS & \cite{Stella1987}  \\ 
4U 0513$-$40         & 11.9    & 17       & P & NS & \cite{Zurek2009}   \\
2S 0918$-$549        & 5.4     & 17.4     & P & NS & \cite{Zhong2011}   \\
4U 1543$-$624        & $\sim6$ & 18.2     & P & NS & \cite{Wang2004,Wang2015}    \\
4U 1850$-$087        & 7.38    & 20.6     & P & NS & \cite{Homer1996}   \\
M15 X-2              & 10.7    & 22.6     & P & NS & \cite{Dieball2005} \\
47Tuc X-9            & 4.5     & 28       & P & BH?& \cite{Bahramian2017}\\
IGR J17062$-$6143    &$\sim7.3$& 38       & T & NS & \cite{Strohmayer2018}\\ 
XTE J1807$-$294      & $\sim8$ & 40.1     & T & NS & \cite{Markwardt2003}\\ 
XTE J1751$-$305      & $\sim8$ & 42       & T & NS & \cite{Markwardt2002}\\ 
4U 1626$-$67         & $\sim8$ & 42       & P & NS & \cite{Chakrabarty1998}\\ 
XTE J0929$-$314      & $\sim8$ & 43.6     & T & NS & \cite{Galloway2002}\\  
MAXI J0911$-$655     & 10      & 44.3     & T & NS & \cite{Sanna2017}   \\  
IGR J16597$-$3704    & 7.2     & 46.0     & T & NS & \cite{Sanna2018}  \\  
4U 1916$-$053        &$\sim9.3$& 50       & P & NS & \cite{Walter1982}  \\
Swift J1756.9$-$2508 & $\sim8$ & 54.7     & T & NS & \cite{Krimm2007}   \\  
NGC 6440 X-2         & 8.2     & 57.3     & T & NS & \cite{Altamirano2010}\\  
IGR J17494$-$3030    & ?       & 75       & T & NS & \cite{Ng2021}\\ 
2FGL J1653.6$-$0159  & $\sim1$ & 75       & Q$^a$ & NS & \cite{Kong2014,Romani2014}\\
\hline
\multicolumn{6}{l}{{\bf Candidate UCXBs}}\\
4U 1728$-$34         & 5.2     & 10.8$^b$ & P & NS & \cite{Galloway2010}\\
4U 0614+091          & 3.2     & 51$^c$   & P & NS & \cite{Shahbaz2008} \\
1A 1246$-$588        & 4.3     & ?        & P & NS & \cite{intZand2008} \\
4U 1812$-$12         & $\sim4$ & ?        & P & NS & \cite{Bassa2006, ArmasPadilla2020}  \\
XMMU J181227.8$-$181234 &$\sim14$& ?      & T & NS & \cite{Goodwin2019}\\
1RXS J180408.9$-$342058 &$<5.8$& ?$^d$    & T & NS & \cite{Baglio2016}\\
IGR J17285$-$2922    & ?       & ?        & T & BH?& \cite{Stoop2021}\\
Swift J0840.7$-$3516 & ?       & ?        & T & ?  & \cite{CotiZelati2021}\\
4U 1857+01           & ?       & ?        & T & NS & \cite{Jonker2006}\\
SAX J1712.6$-$3739   & $\sim7$ & ?        & T & NS & \cite{intZand2007}\\
1RXS J170854.4$-$321857&$\sim13$& ?       & P & NS & \cite{intZand2007,ArmasPadilla2019}\\ 
1RXS J171824.2$-$402934& ?       & ?        & P & NS & \cite{intZand2009}\\
4U 1722$-$30           & 7.7     & ?        & P & NS & \cite{intZand2007}\\
1RXS J172525.2$-$325717& ?       & ?        & P & NS & \cite{intZand2007}\\
SLX 1735-269           & $\sim6$ & ?        & P & NS & \cite{intZand2007}\\
SLX 1737-282           & 7.3     & ?        & P & NS & \cite{intZand2007}\\
SLX 1744-299           & ?       & ?        & P & NS & \cite{intZand2007}\\
XMMU J174716.1-281048  &$\leq8.4$& ?        & P & NS & \cite{degenaar2011burst}\\
 
\hline
\end{tabular}
\caption{A catalog of known Galactic UCXBs and candidates. Table partially based on \cite{intZand2007} and \cite{Cartwright2013}. ``T/P'' indicates whether the source is transient (T) or persistent (P). ``NS/BH'' indicates nature of the compact object (e.g., determined via detection of Type I X-ray bursts or pulsations). Currently, there are no confirmed BH-UCXB systems identified. Candidate UCXBs are generally identified based on their characteristics suc as signatures of hydrogen deficiency (e.g., long Type I X-ray bursts, featureless optical spectrum, or prominence of neon, carbon or oxygen features). a- 2FGL~J1653.6$-$0159 was identified as a Gamma ray millisecond pulsar with a quiescent (L$_X\sim10^{31}$ \ergs) X-ray counterpart. b- \cite{Galloway2010} identify a suggestive periodic signal at 10.8 min, however we note that \cite{Vincentelli2020} recently suggest that the orbital period is unlikely to be shorter than 66 min and thus the system is not a confirmed UCXB. c- An orbital period of $\sim50$ has been suggested by \cite{Shahbaz2008}; \cite{Baglio2014} suggest that the orbital period may be $>1$ hour. However, the authors note that the companion still appears to be hydrogen-deficient (thus still likely to be a UCXB). d- An orbital period of 40 minutes is indirectly inferred by \cite{Baglio2016} based on evolutionary tracks of a presumed He donor. However, this periodicity is not yet observationally confirmed, specially as \cite{Marino2019} note possible presence of H/He in the accreted material.}
\label{tab:UCXBs}
\end{table}

\subsection{Eclipsing LMXBs}\label{subsec:eclipsing}
Eclipsing LMXBs are not a characteristically separate class of LMXBs, as the eclipses are merely a result of high inclination angle ($i\sim90^\circ$) of the binary orbital plane from our point of view (see Figure~\ref{fig:binsim}). However, our observational perspective of eclipsing binaries provides a unique direct method to study properties of accretion disk via eclipse mapping \citep{Baptista2001}, and evolution of orbital period in LMXBs on timescales of years/decades \citep[e.g.,][]{Chou2014}. It has also been suggested that study of eclipses in accreting systems in the radio band can prove powerful in understanding jet physics \citep{maccarone2020_eclipse}.

\begin{figure}
    \centering
    \includegraphics[scale=0.55]{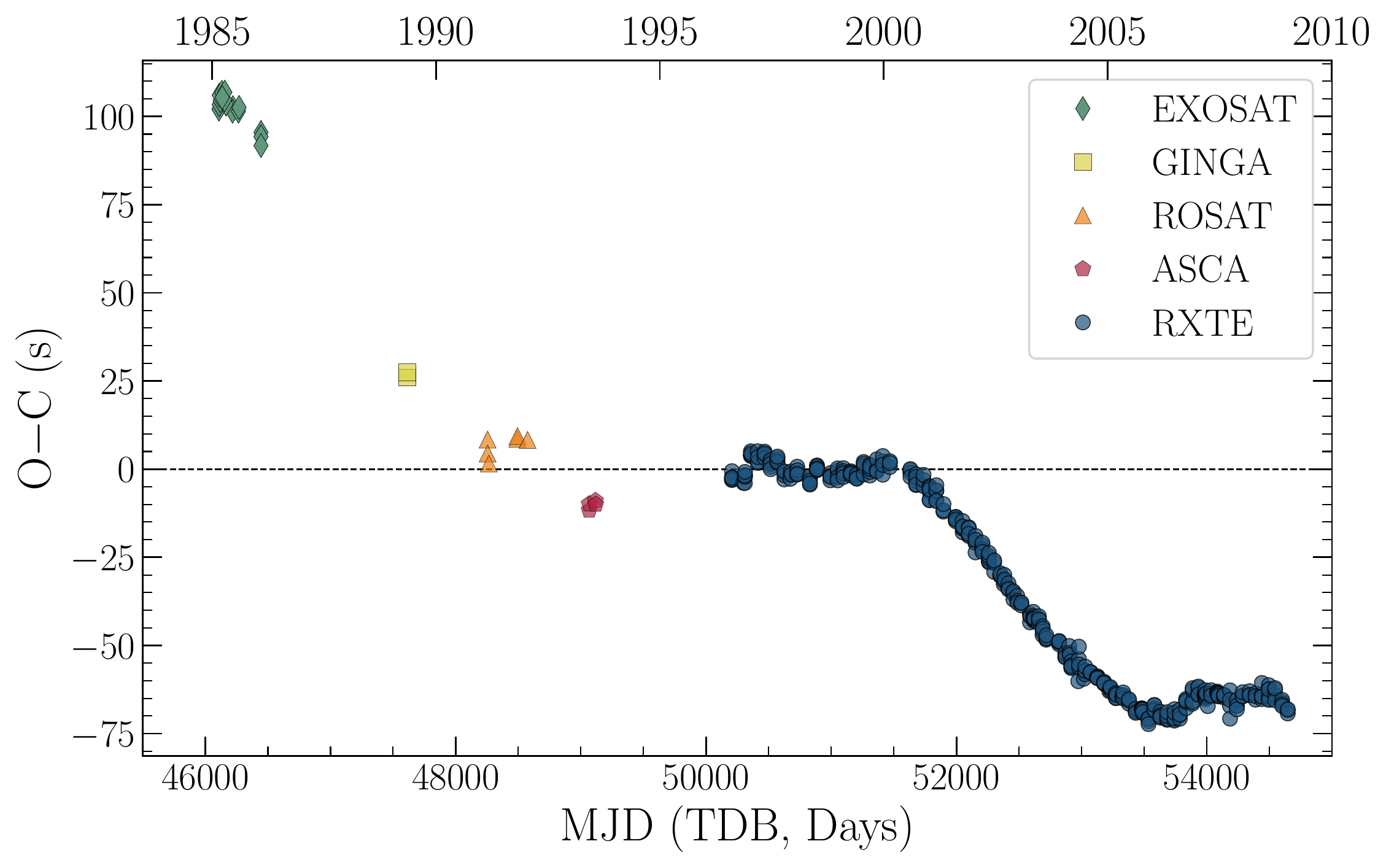}
    \caption{Evolution of orbital period in EXO 0748-676 (based on \citep{Wolff2009}). The change in orbital period does not follow a single pattern and appears to show epochs, with different rates of change. EXO 0748-676 is among the best documented cases of such anomalous orbital changes in LMXBs, however it is not the only eclipsing LMXB exhibiting irregular variations in orbital period evolution.\\}
    \label{fig:eclipse}
\end{figure}

Broadly speaking, the orbital period in X-ray binaries is expected to decay, as momentum is lost in accretion processes, outflows, gravitational radiation, and magnetic breaking \citep[e.g.,][]{Paczynski1967, Verbunt1981, Tavani1991}. Studying the evolution of orbital period on timescale of years/decades in HMXBs has shown that such a decay is indeed present and can be generally described by a smooth linear or quadratic ephemeris model \citep[e.g.,][]{Falanga2015}. However, observations of LMXBs indicate while many of them show orbital decay, the changes in the orbital decay in some cases cannot be well described by linear or quadratic ephemeris models \citep{Jain2011,Wolff2009, Iaria2018, Ponti2017}, and many of these systems exhibit multiple ``epochs'' of orbital decay with highly variable rates (Figure~\ref{fig:eclipse}), sometimes accompanied by period ``jitters''. These anomalies demonstrate that evolution of orbital period in LMXBs on short timescales (years) is not fully understood.

Currently there are $\sim$a dozen eclipsing Galactic LMXBs identified (Table~\ref{tab:eclipsingLMXBs}). Almost all of these systems have been identified as eclipsing LMXBs during bright outbursts. However, a select few have been identified in quiescence in deep observations (e.g., X5 in globular cluster 47 Tuc, which has been observed substantially by the \chandra\ X-ray observatory). In addition to the sources cataloged in Table~\ref{tab:eclipsingLMXBs}, there are a handful of other eclipsing systems that either show only partial eclipses \citep[e.g., see ][and references therein]{Chou2014}, or while eclipses have been observed period or nature of the system are uncertain \citep[e.g.,][]{Maeda2013}. 

\begin{table}
\centering
\begin{tabular}{llccc}
\hline
\hline
System                  & $P_\textrm{orb}$  &$\dot{P}_\textrm{orb}$     & NS/BH &   Reference \\
                        & (hour)            & ($10^{-12}$ s s$^{-1}$)   &       &   \\
\hline
XTE~J1710$-$281$^a$     & 3.281063218(7)    &$-1.6\leq\dot{P}_\textrm{orb}\leq0.2$& NS & \citep{Jain2011} \\
EXO~0748$-$676$^b$      & 3.824088(1)       &     -                     & NS    & \citep{Wolff2009}   \\
4U~2129+47$^c$          & 5.238220(2)       &$103(\pm13)$               & NS    & \citep{Bozzo2007}   \\
IGR~J17451$-$3022       & 6.2834(5)         &     -                     & ?     & \citep{Bozzo2016}   \\
H~1658$-$298$^d$        & 7.1161099(3)      &$-8.5(\pm1.2)$             & NS    & \citep{Wachter2000, Iaria2018} \\
CXOGC~J174540.0$-$290031& 7.767(2)          &     -                     & ?     & \citep{Muno2005a,porquet2005}\\
AX~J1745.6$-$2901       & 8.3510081(2)      &$-40.3(\pm2.7)$            & ?     & \citep{Ponti2017}    \\
47~Tuc~X5               & 8.67(1)           &     -                     & NS    & \citep{Heinke2003}   \\
3FGL~J0427.9$-$6704     & 8.80128(2)        &     -                     & NS    & \citep{Strader2016}  \\
Swift~J1749.4$-$2807    & 8.816866(2)       &     -                     & NS    & \citep{Markwardt2010}\\
GRS~J1747$-$312         & 12.3595273(2)     &     -                     & NS    & \citep{intZand2003}  \\
Swift J1858.6$-$0814    & 21.3448(4)        &     -                     & NS    & \cite{Buisson2021}   \\
Her~X-1$^e$             & 40.80402216(5)    &$-48.5(\pm1.3)$            & NS    & \citep{Staubert2009} \\
\hline
\end{tabular}
\caption{A catalog of known Galactic LMXBs exhibiting clear total eclipses (for the purpose of this catalog we omit LMXBs with suggestive or partial eclipses). ``NS/BH'' indicates nature of the compact object (via detection of Type I X-ray bursts or pulsations, or deep X-ray spectroscopy in case of quiescent systems like 47~Tuc~X5). No eclipsing BH-LMXB has been confirmed as of this writing. $^a$- \citep{Jain2011} demonstrate that XTE J1710-281 shows irregular variations, and find the best-fit ephemeris solution is a piece-wise linear model. $^b$- EXO 0748-676 shows strong variations in ephemeris and \citep{Wolff2009} indicate no simple linear or quadratic solution describes all data. $^c$- \citep{Bozzo2007} speculate that the ephemeris of 4U 2129+47 indicates that it could be in fact part of a hierarchical triplet system. $^d$- The best-fit ephemeris solution suggested by \citep{Iaria2018} contained linear, quadratic and a sinusoidal term with a period of 2.31 yr. The authors suggest such a modulation could be produced by the gravitation coupling of the orbit with changes in the shape of the magnetically  active companion star. $^e$- We note that Her X-1 is not strictly a LMXB as the companion is estimated to be $\sim 2$M$_\odot$.} 
\label{tab:eclipsingLMXBs}
\end{table}

\subsection{Wind-fed accretion in LMXBs: Symbiotic X-ray binaries}\label{subsec:symbiotics}
In contrast with other classes of LMXBs, symbiotic X-ray binaries (SyXBs) stand out by the absence of Roche-lobe overflow. These systems consist of a compact object (BH or NS) accreting from the winds of a low-mass late-type giant. It is important to clarify that we define symbiotic X-ray binaries distinctly from ``symbiotic stars'' (or symbiotic binaries), which are defined by a white dwarf accreting from winds of a late-type giant.\footnote{In classification of symbiotic stars and systems based on their X-ray properties (e.g., the groups labeled as $\alpha$, $\beta$, $\gamma$, $\delta$, $\beta/\delta$), symbiotic X-ray binaries are generally classified as the $\gamma$ group \citep{Murset1997, Luna2013}.}

GX 1+4 is the only SyXBs identified so far that has been observed to be persistently bright ($\geq 10^{36}$ \ergs) in the X-rays \citep{Lewin1971, Serim2017, Ilkiewicz2017}. However, most SyXBs show strong rapid variability, with quiescence luminosity $L_X \leq 10^{33}$ \ergs\ and outburst $L_X \geq 10^{34}$ \ergs\ within hours/days, and can decay as quickly \citep[e.g.,][]{Patel2007, Masetti2007, Heinke2009b}. Additionally, SyXBs tend to show rapid ``faint'' outbursts with peak luminosity $10^{34}$ \ergs\ $\leq L_X \leq 10^{36}$ \ergs\ \citep{Masetti2006, Kaplan2007, Farrell2010, Kuranov2015}. The sporadic nature of their variability makes SyXBs particularly difficult to detect and identify, thus characterizing their population and behavior requires sensitive frequent monitoring \citep[e.g., with \swift/XRT or {\it eROSITA};][]{Bahramian2021}. 

To date, no BH SyXB system has been identified. In a large fraction of SyXBs, discovery of spin periods (in the 100s to 1000s of seconds range) have led to identification of the NS accretor \citep[][see also Table~\ref{tab:SyXBs}]{Patel2004, Bodaghee2006, Thompson2006, Corbel2008}. Companion stars in SyXBs are typically K-M III giants \citep{Chakrabarty1997, Masetti2002, Bozzo2013, Bahramian2014a, Shaw2020}. However, in rare cases the companions have been identified as carbon stars \citep{Masetti2011, Hynes2014}. Owing to the detached nature of SyXBs and the size of the companion, these systems typically have long orbital periods ranging from tens to thousands of days and in many cases difficult to constrain with current available coverage of these systems \citep{Hinkle2006, Nespoli2010, Kuranov2015}.

\begin{figure}
    \centering
    \includegraphics[scale=0.55]{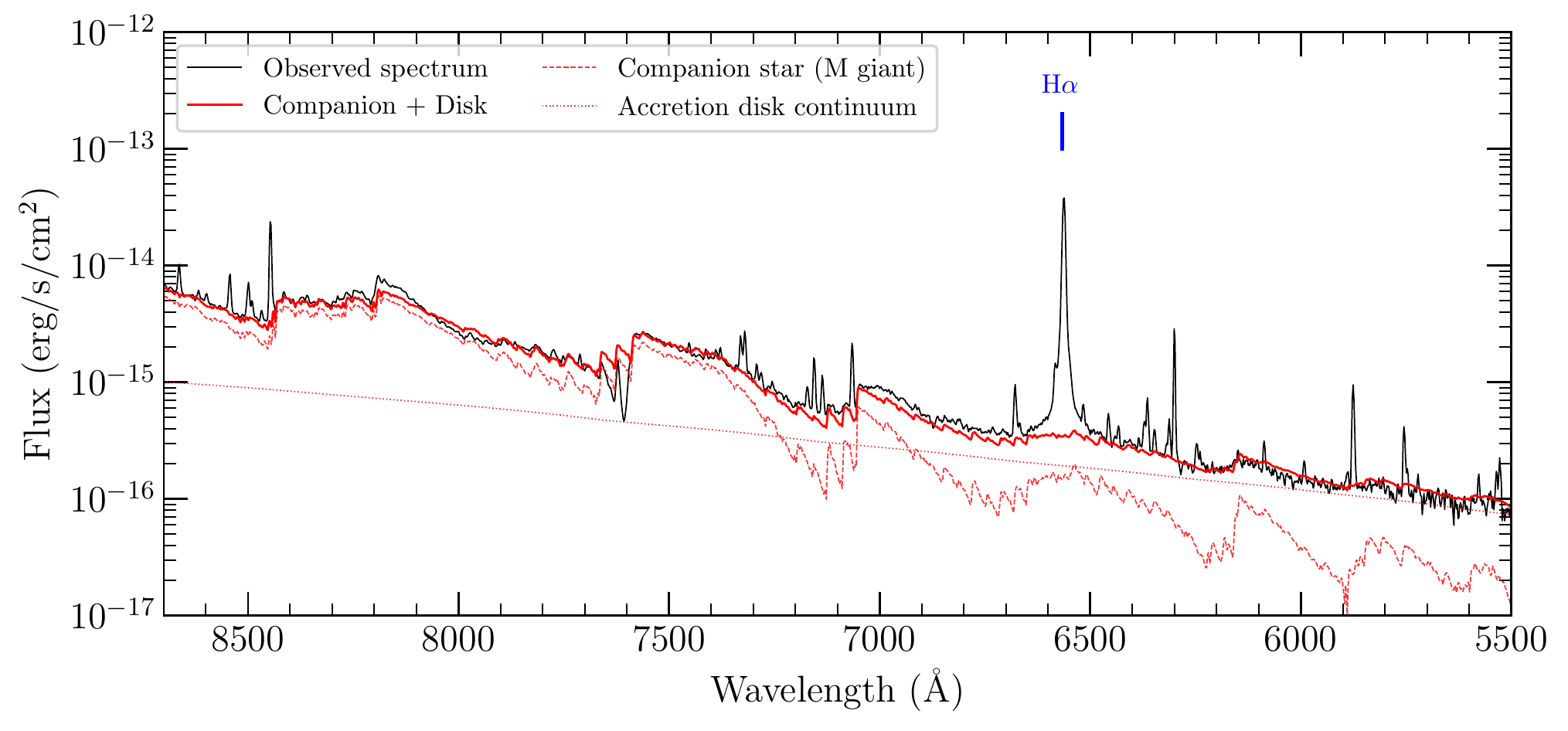}
    \caption{Optical spectrum of a typical Symbiotic X-ray binary (IGR J17329-2731) during an outburst. While generally, the optical and infrared continua are dominated by the companion, strong accretion features like broad or double-peaked H-$\alpha$ lines can sometimes be visible. During outbursts, the continuum emission from the accretion disk (i.e., multi-color blackbody) can also be observed in some cases at shorter wavelengths. Figure based on \citep{Bozzo2018}.}
    \label{fig:syxb}
\end{figure}

The spectrum of SyXBs in the optical and NIR regimes is typically dominated (almost completely) by the companion star. In some cases, strong hydrogen emission lines like H-$\alpha$ or Br-$\gamma$ (sometimes broad and/or double-peaked) are also observed, particularly during enhanced X-ray activity \citep{Chakrabarty1997, Nespoli2010}. Furthermore, during enhanced activity, contribution from the accretion continuum can be noticed at shorter wavelengths \citep[Fig.~\ref{fig:syxb};][]{Bozzo2018}.

Study of accretion and outflows in SyXBs in the radio frequencies is also challenging due to their rapid variability. However, radio observations of GX 1+4 - the only persistent SyXB, has allowed study of these mechanisms in SyXBs. While earlier radio observations of this system reported non-detections or marginal detections \citep{Seaquist1993, Fender1997, Marti1997}, later observations by the Karl G. Jansky Very Large Array showed significant detection of emission, which perhaps could be caused by the outflow \citep{vandeneijnden2018b}.

\begin{table}
\centering
\begin{tabular}{lcccccc}
\hline
\hline
System              & $P_{\textrm{spin}}$ & $P_{\textrm{orbit}}$  &   Peak L$_x$      &   Distance    &  NS/BH & Reference  \\
                    & s                        & d                          &   \ergs           &   kpc         &    &            \\
\hline
GX~1+4              & 140                      & 1161                       &$10^{36}$          & 4.3           & NS & \citep{Chakrabarty1997,Hinkle2006,Ferrigno2007,GonzalezGalan2012}\\
Sct~X-1             & 113                      & ?                          &$2\times10^{34}$   & $\geq 4$      & NS & \cite{Kaplan2007}\\
3XMM J181923.7–170616 & 408                    & ?                          & ?                 & ?             & NS & \citep{Qiu2017}\\
IGR~J16358$-$4726   & 5850                     & ?                          &$3\times10^{36}$   & 5--13         & NS & \citep{Patel2004,Patel2007,Nespoli2010,Lutovinov2005}\\
IGR~J17329$-$2731   & 6680                     & ?                          &$3\times10^{35}$   & $\sim2.7$     & NS & \citep{Bozzo2018}\\
CGCS~5926           & ?                        & $\sim$151                  &$3\times10^{32}$   & $6(\pm1)^a$   & ?  & \cite{Masetti2011}\\
4U~1700+24          & ?                        & 4391$^c$                   &$10^{34}$          & 0.544         & NS & \cite{Masetti2002,Masetti2006,Hinkle2019} \\
IGR~J16194$-$2810   & ?                        & ?                          &$\leq10^{35}$      &$2.1(\pm0.2)^a$& ?  & \cite{Masetti2007}\\
CXOGBS~J173620.2$-$293338 & ?                  & ?                          &$10^{33}$          & $\sim8^{a,b}$ & ?  & \citep{Hynes2014}\\
XTE~J1743$-$363     & ?                        & ?                          & ?                 & $\sim8^{a,b}$ & ?   & \citep{Smith2012,Bozzo2013}\\
IGR~J17445$-$2747   & ?                        & ?                          & ?                 & 1.1--7.6      & NS & \citep{Mereminskiy2017,Shaw2020}\\
XMMU~J174445.5$-$295044 & ?                    & ?                          & $10^{35}$         & 3.1           & ?  & \citep{Bahramian2014a}\\
IGR~J17197$-$3010   & ?                        & ?                          &$\leq2\times10^{35}$& 6--17        & ?  & \citep{Masetti2012b}\\
CXOGBS~J174614.3$-$321949 & ?                  & ?                          & $10^{34}$         & $\sim8^{a,b}$ & ?  & \citep{Wetuski2021}\\
CXOGBS~J173620.2$-$293338 & ?                  & ?                          & $\leq10^{33}$     & $\sim8^{a,b}$ & ?  & \citep{Wetuski2021}\\
IGR~J17597$-$2201   & ?                        & ?                          & ?                 & $15$?         & NS & \citep{Chaty2008,Ratti2010,Zolotukhin2015}\\
Swift~J2037.2+4151  & ?                        & ?                          & $10^{36}$         & $\sim10$      & ?  & \citep{Molina2021}\\
\hline
\end{tabular}
\caption{A catalog of identified Galactic SyXBs and SyXB candidates. Table partially based on \cite{Yungelson2019}, see also \cite{Kuranov2015, Merc2019}. ``NS/BH'' indicates nature of the compact object (via detection of Type I X-ray bursts or pulsations). No Symbiotic BH-LMXB has been confirmed as of this writing. It is worth noting that Peak L$_X$ is the \emph{reported} peak luminosity; SyXBs can have short lived flares (minutes long) reaching higher luminosities that may have been missed in pointed observations too short to be noticed in X-ray monitors. SyXB candidates not listed in this table include 4U~1954+31 (proposed to be a SyXB by \cite{Corbel2008, Masetti2006}, recent works indicate it is likely an HMXB \citep{Hinkle2020}), IRXS~J180431.1$-$273932 (suggested as SyXB by \citep{Nucita2007}, disputed by \citep{Masetti2012}), IGR~J16393$-$4643 (suggested as a SyXB by \citep{Nespoli2010}, disputed by \citep{Bodaghee2012b}), 2XMM~J174016.0$-$290337 (suggested as a SyXB by \cite{Farrell2010} and disputed by \citep{Thorstensen2013}), Swift~J175233.3$-$293944 (suggested as symbiotic by \citep{Wevers2017}, a white dwarf compact object is favored by \citep{Bahramian2021}).\\ $^a$ - indicates the distance (and luminosity) has been updated based on Gaia EDR3 \cite{GaiaEDR3summary2021}. This includes applying parallax zero-point correction based on \citep{Lindegren2021} and considering a Galactic prior based on distribution of LMXBs \citep{Atri2019}. $^b$ - Indicates that while a parallax value is reported EDR3 for the source, it is not significant and thus distance estimation is dominated by prior assumptions. $^c$- An alternative period of 404 days is also discussed by \citep{Galloway2002, Hinkle2019}.}
\label{tab:SyXBs}
\end{table}

\subsection{Magnetically-channeled accretion in LMXBs: X-ray pulsars}\label{subsec:pulsars}

In a sub-group of NS-LMXBs, material from the inner part of the accretion disk is funneled along the magnetic field lines of the NS onto its magnetic poles. This localized accretion causes hotspots that give rise to X-ray pulsations modulated at the spin period of the NS \cite{Bildsten1997,Wijnands1998}. Roughly 20\% of NS-LMXBs display rapid X-ray pulsations, in a range of $\sim$1--10~ms, and are called Accreting Millisecond X-ray Pulsars (AMXPs; see \cite{campana2018} for an overview until 2018 and \cite{Sanna2018,Sanna2018b,Ng2021} for 3 new AMXPs discovered since). The NSs in AMXPs have relatively weak magnetic field strengths of $B < 10^{9}$~G \cite{Mukherjee2015}. Furthermore, these NSs are thought to have acquired their rapid spin by angular momentum transfer in the accretion process, and are likely the progenitors of millisecond radio pulsars (MSRPs) \cite{Bhattacharya1991,Alpar1982}. 

The connection between AMXPs and MSRPs seems to be supported by the discovery of the so-called ``transitional millisecond radio pulsars'' (tMSRPs); these are NSs in binary systems that are sometimes observable as regular millisecond radio pulsars, but at other times look more like a LMXB with an accretion disk present and no observable radio pulsations \cite{campana2018}. Currently, there are 3 systems showing clear evidence for switches between these two manifestations, which occur on a timescale of years (IGR J18245--2452 in the globular cluster M28, PSR J1023+0038 and XSS J12270--4859; \cite{Archibald2009,Papitto2013,Bassa2014}). Given the scientific interest of studying systems that appear to bridge different NS populations, there has been much effort in finding more tMSRPs. This has resulted in several good candidate systems (see \cite{papitto2020} for a recent overview), but so far no new confirmed ones have emerged yet. 

The binary parameters of AMXPs (and tMSRPs) are well known through accurate timing of their (X-ray) pulsations. In a handful of systems, for which the orbital period is $\sim$2~hr, the companion is thought to be a brown dwarf. The remaining AMXPs are split into roughly equal groups of systems with white dwarf or helium donor stars in tight orbits ($\sim$0.25--1.25~hr, i.e., UCXBs), and systems with main sequence donor stars and longer orbital periods (nearly all $\sim$3--11~hr) \cite{campana2018}. The three confirmed tMSRPs all fall in the latter group.  

Apart from the rapidly spinning AMXPs, a very small subgroup ($\lesssim$10\%) of NS LMXBs displays much slower pulsations (seconds to hours). In addition to slower spin, the NSs in these systems also have stronger magnetic fields ($B \sim 10^{10-12}$~G) than the rapidly spinning AMXPs. There are 4 slow pulsars in Roche-lobe overflow LMXBs (2A 1822--371, 4U 1626--67, GRO 1744--28, and Her X-1)\footnote{Her X-1 formally classifies as an intermediate-mass X-ray binary (IMXB).} and their spin periods are in the range of $\sim$0.1--2~s.  In addition, there are 5 slow pulsars found in the wind-fed SyXBs which have spin periods of $\sim$10~min to $\sim$5~hr (see Table~\ref{tab:SyXBs}). 

For the wind-accreting systems, the accretion efficiency is likely too low to transfer significant amounts of angular momentum that spin up the NS. However, for the slow pulsars in Roche-lobe overflow systems, it is not yet clear why these NSs have not been spinned up. The explanation likely lies in the prior evolution of the binary \cite{verbunt1990,Rappaport1997,vanparadijs1997}. In terms of binary parameters, the slow LMXB pulsars are a heterogeneous group with orbital periods from $<$1 hr to $>$11 days and white dwarf, main-sequence or giant companions.


\section{Variability and transient outbursts in LMXBs}\label{sec:outbursts}
In addition to the physical classification based on compact object type, donor type and mode of mass transfer discussed in the previous sections, LMXBs can be phenomenologically categorized based on their X-ray behavior. In the following sections we provide an overview of classifications based on their long-term X-ray variability (on $\sim$ months/year timescales), short-term X-ray variability (on $\sim$ days/weeks timescales), and X-ray (peak) luminosity. As we will see, the dynamic range in X-ray luminosity traced out by the population of LMXBs as a whole covers many orders of magnitude; from $\lx \simeq 10^{39}$~\ergs\ for sources accreting around the Eddington luminosity to $\lx \simeq 10^{30}$~\ergs\ for quiescent systems in which little or no accretion occurs.

It is important to remind the reader that with these various ways to (physically or phenomenologically) classify LMXBs, there is overlap between different categories. For instance, several AMXPs are UCXBs and the broad distinctions of persistent and transient sources contain basically all other sub-classes. Furthermore, NS-LMXBs can be divided into different categories based on their X-ray properties (AMXPs, slow pulsars, non-pulsating systems, atolls, Z-sources), as briefly discussed in Section~\ref{subsec:lum_class}.

\subsection{Long-term X-ray behavior: Transient and persistent LMXBs}\label{subsec:trans_pers}
\begin{figure}
    \centering
    \includegraphics[scale=0.5]{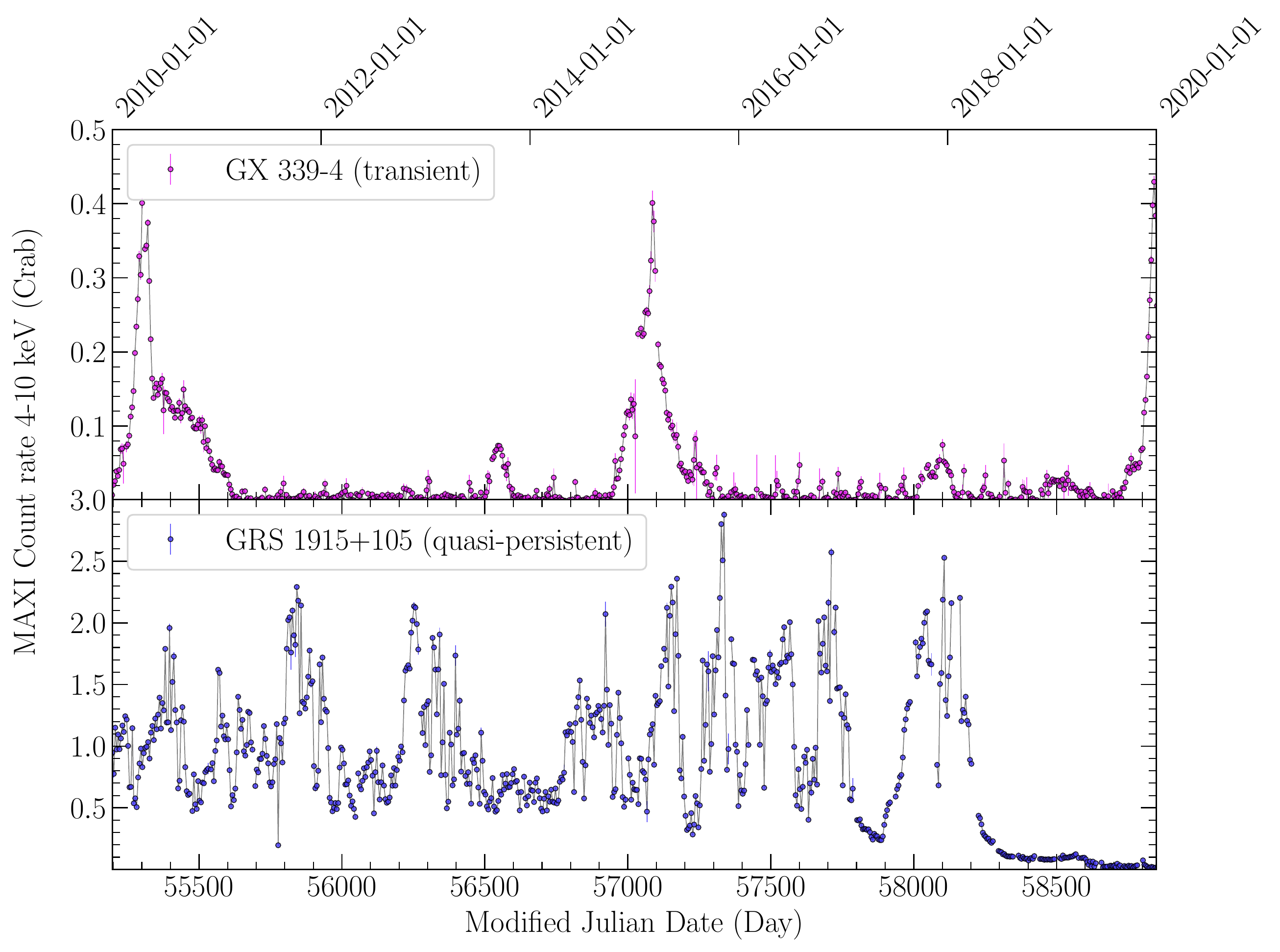}
    \caption{Light curves of LMXBs and their variability over the past decade based on monitoring data from \maxi/GSC  \citep{Matsuoka2009}. Top: The transient black hole LMXB GX339-4, showing several distinct accretion outbursts of different brightness. Bottom: The quasi-persistent black hole LMXB GRS~1915+105, illustrating that large variations in brightness can also occur in systems that are continuously active.}
    \label{fig:lcs}
\end{figure}

One of the primary classifications of LMXBs is based on their long-term X-ray behavior, which separates the persistent systems from the transient ones. Persistent LMXBs are always actively accreting, but can still show (in some cases quite strong) X-ray variability (e.g. Figure~\ref{fig:lcs} bottom). Transient LMXBs, on the other hand, exhibit large swings in their X-ray luminosity that separate outbursts of active accretion with periods of quiescence (an example is shown in Figure~\ref{fig:lcs} top).  

Whereas there is no strict definition of what level of X-ray luminosity variability defines a transient, generally LMXBs are considered transient if their X-ray luminosity changes by a factor of $\gtrsim$1000. The long-term light curves shown in Figure~\ref{fig:lcs} are examples of the differences in long-term X-ray behavior observed among LMXBs. These light curves are based on publicly available data from the all-sky X-ray monitoring of 
\maxi\ (2--20~keV; \cite{Matsuoka2009}). 

\subsubsection{Extended outbursts: Quasi-persistent LMXBs}\label{subsubsec:qp}
Typically, the outbursts of transient LMXBs last for a few weeks up to $\sim$a year, while their inactive quiescent phases may extend for many years or even decades. A more detailed overview of this follows in Section~\ref{subsubsec:outbursts}, but here we note that a small sub-class of LMXBs falls more or less in between the two categories defined above: these systems are formally transient (i.e., clearly switching between active and inactive periods), but exhibit extended outburst episodes that last for years or even decades. An extreme example is the (eclipsing) NS-LMXB EXO~0748--676, which exhibited an outburst of 24 years between the early 80s and 2009 \cite{Degenaar2011}. Systems with extended outbursts of $\gtrsim1$~yr have been dubbed ``quasi-persistent'' LMXBs. 

It is worth noting that the quasi-persistent LMXBs include a handful of systems that were observed to switch on at some point in time, but have never returned to quiescence since. Examples are the BH-LMXB GRS 1915+105, which has been active since 1992 (e.g., \cite{Motta2021}), and the NS system IGR J17062--6143, which has been accreting actively since 2006 \cite{Hernandez2019}. On the other hand, there are also LMXBs that were once thought to be persistent because they continued to be detected for years following their original discovery, but then suddenly dimmed into quiescence. One example is the NS-LMXB X1732--304 in the globular cluster Terzan 1, which was discovered in 1980 and suddenly disappeared in the late 90s \cite{Guainazzi1999}. 
The cause of the extended outbursts is not fully clear, but may lie in mass-transfer variations from the donor star, e.g., due to irradiation or moving Sun spots \cite{King1998b,Shaw2019}.\footnote{The standstill observed for Z Cam stars \cite{hameury2020} could be the white-dwarf analogues of the extended outbursts of quasi-persistent LMXBs \cite{Shaw2019}.} 
Table~\ref{tab:QP} lists LMXBs known to display quasi-persistent outbursts. 

\begin{table}
\centering
\begin{tabular}{lccc}
\hline
\hline
System              & current state  & sub-class & Reference \\
\hline
GRS 1915+105     &    active$^{a}$     &     BH                 &  \cite{Miller2020,Motta2021}\\
Swift J1753.5--0127     &    quiescent     &     BH                 & \cite{Zhang2019} \\
IGR J17098--3628  &  quiescent       &     BHC                & \cite{Capitanio2009} \\   
IGR J17091-3624 &  quiescent       &     BHC                & \cite{Pereyra2020} \\
EXO 0748--676     &    quiescent     &     NS, eclipsing                 & \cite{Degenaar2011} \\
MXB 1659--298     &    quiescent     &     NS, eclipsing                 & \cite{Wijnands2003} \\
KS 1731--260     & quiescent &      NS                  & \cite{Wijnands2001} \\
HETE J1900.1--2455     & quiescent &      NS, AMXP                  & \cite{degenaar2017_hete} \\
MAXI J0556--332$^{c}$     & quiescent &     NS, transient Z-source                 & \cite{Homan2014} \\
XTE J1701--462     & quiescent &      NS, transient Z-source                 & \cite{Fridriksson2010} \\
XTE J1701--407     & active$^{b}$ &      NS                 & \cite{degenaar2011_qp} \\
IGR J17062--6143    & active$^{b}$ &     NS, VFXB, AMXP                 & \cite{Hernandez2019} \\
Swift J0911.9--6452 (NGC 2808)    & active &     NS, AMXP, UCXB                & \cite{Ng2021_0911,Sanna2017} \\
Swift J1858.6--0814     & quiescent &     NS, accreting near Eddington   & \cite{Buisson2020,parikh2020_sw} \\
1M 1716--315     & quiescent &     NS                 & \cite{jonker2007qp} \\
1H 1905+000     &   quiescent      &     NS                 & \cite{Jonker2006} \\
AX J1745.6--2901$^{c}$ (Galactic center)          & active &     NS, eclipsing               & \cite{Degenaar2015GC} \\
AX J1754.2--2754     &    active$^{b}$     &     NS, VFXB                 & \cite{degenaar2012asca} \\
XMMU J174716.1--281048  &    quiescent     &     NS, VFXB                 & \cite{DelSanto2007}\\
4U 2129+47 &     quiescent    &     NS                 & \cite{Nowak2002} \\   
2S 1711--339  &   quiescent      &     NS                 & \cite{Swank2001} \\ 
XB 1733--30 (Terzan 1)  &   quiescent      &     NS                & \cite{Guainazzi1999} \\   
\hline
\end{tabular}
\caption{List of (confirmed) LMXBs that are quasi-persistent, i.e., transient systems but exhibiting prolonged outbursts of $>$1~yr. 
{\it Notes:}\\ 
$^{a}$ The X-ray emission from GRS 1915+105 suddenly dropped in 2018 and has remained low since, but this is likely caused by local obscuration and not by ceasing accretion activity \cite{Motta2021}. \\
$^{b}$ These (NS) LMXBs accrete at relatively low X-ray luminosity and are therefore not detected by all-sky X-ray monitors such as \swift/BAT and \maxi. Therefore, the status of these sources can only be verified via pointed X-ray observations (e.g., with \swift/XRT); since these generally do not occur regularly, the status of these sources is not accurately known at present. \\
$^{c}$ These quasi-persistent LMXBs are known to also exhibit regular (i.e., shorter) outbursts.}
\label{tab:QP}
\end{table}


\subsubsection{Outburst statistics of transient LMXBs}\label{subsubsec:outbursts}
Systematically studying samples of transient LMXBs has provided insight into their luminosity distribution, outburst duration and duty cycles \cite{Chen1997,Yan2015,Lin2019,Tetarenko2016a}. Many of these studies made use of \rxte, which provided more regular and much denser coverage than any of the earlier X-ray satellites, whilst providing a baseline longer than that of current X-ray monitoring missions (15 years).

\begin{figure}
    \centering
    \includegraphics[width=12cm]{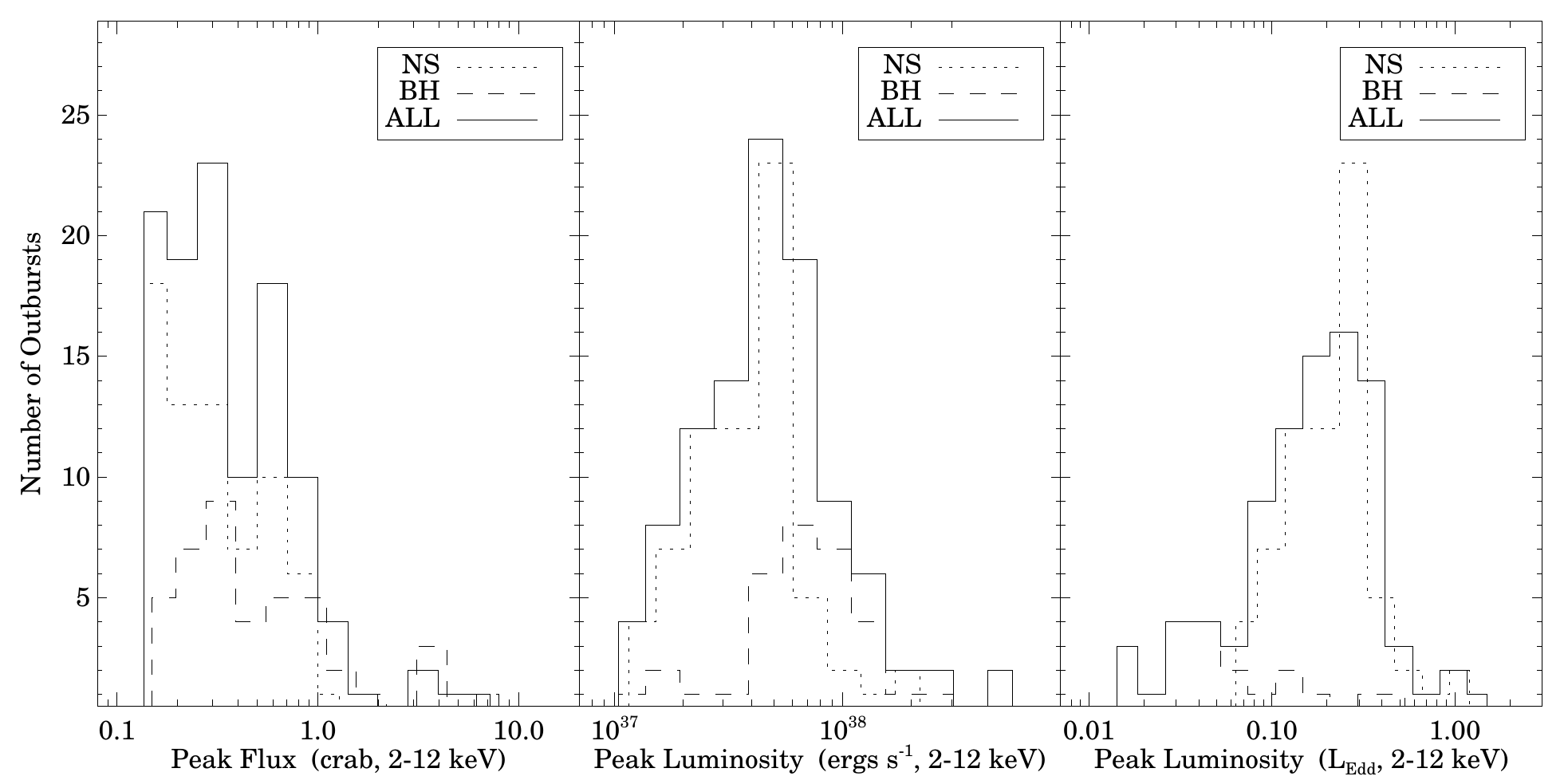}
    \caption{Peak X-ray luminosity distribution of NS and BH LMXB outbursts observed with RXTE between 1996 and 2011 (figure from \cite{Yan2015}).
    }
    \label{fig:Lxpeak_RXTE}
\end{figure}

\begin{figure}
    \centering
    \includegraphics[width=5.8cm]{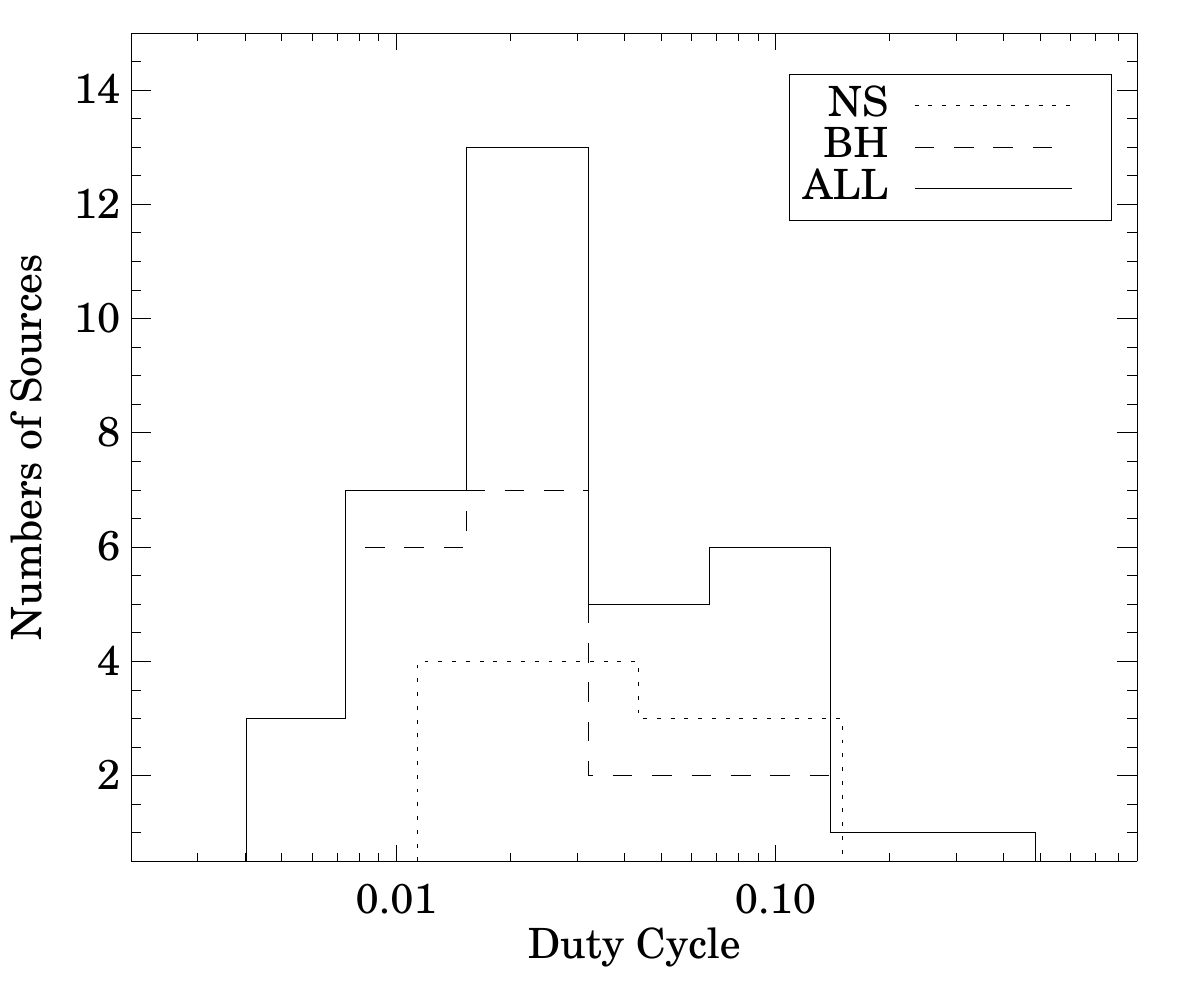}
        \includegraphics[width=5.8cm]{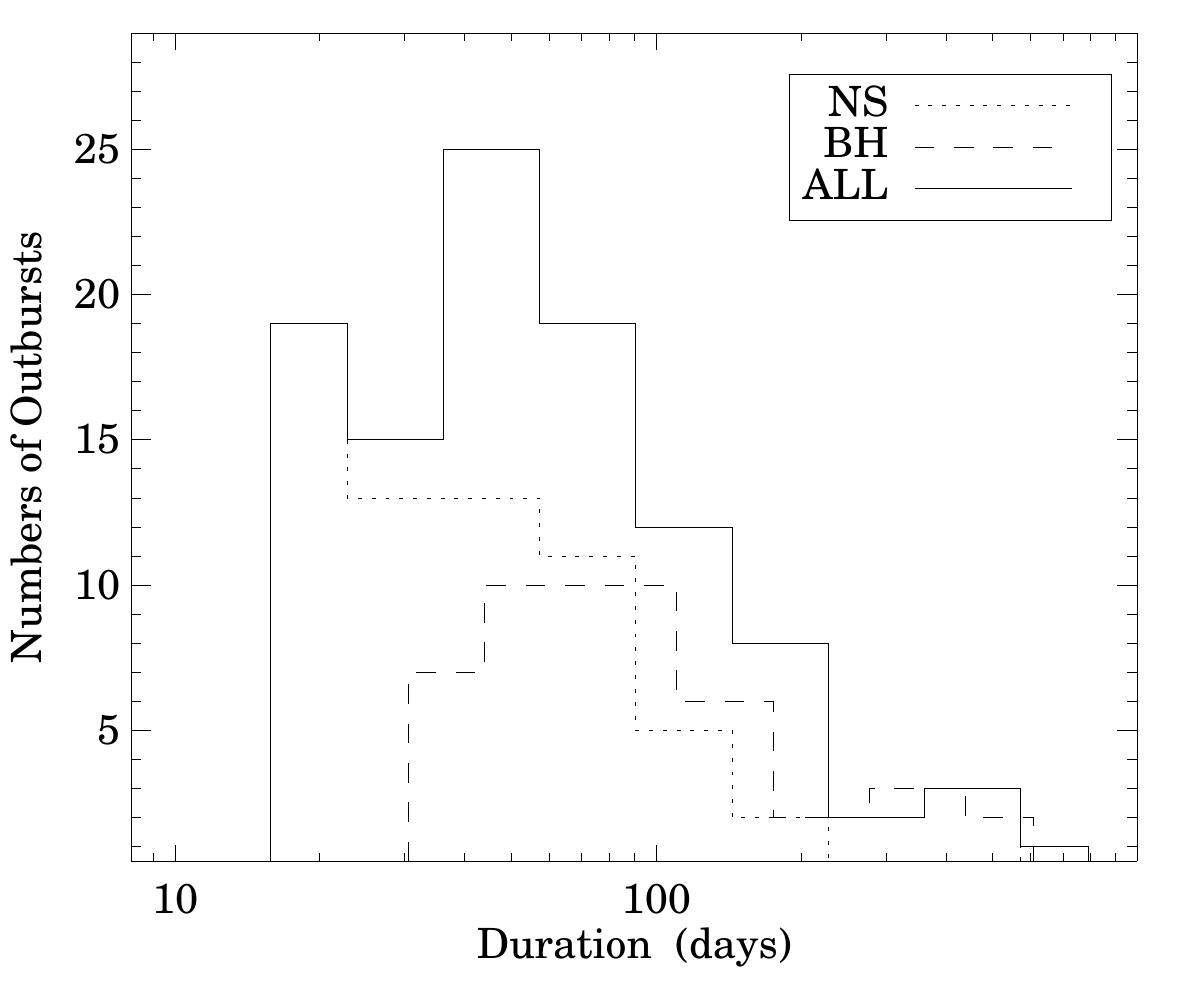}
    \caption{Distribution of the duty cycle (left) and duration (right) of outbursts observed from transient NS and BH LMXBs with \rxte\ between 1996 and 2011 (figure from \cite{Yan2015}).}
    \label{fig:dc_tob_RXTE}
\end{figure}

Yan \& Yu 2015 \cite{Yan2015} performed a systematic analysis of 110 transient outbursts from 36 sources, consisting of 22 (c)BHs and 14 NSs, observed with \rxte\ between 1996 and 2011. This study was limited to fluxes above $\approx 2 \times 10^{-9}$~\ergcms\ (2--12 keV), which translates into a luminosity of $\lx \approx 2\times 10^{37}$~($D$/8~kpc)$^{2}$~\ergs\ at the distance of the Galactic center. The full range of (peak) X-ray luminosities covered by this sample ranged from $\lx \approx 2\times 10^{37}$ to $ 3\times 10^{38}$~\ergs, with an average peak luminosity of $\lx \approx 5\times 10^{37}$~\ergs. This distribution is shown in Figure~\ref{fig:Lxpeak_RXTE}. Furthermore, the range in outburst duration in this sample spanned $\approx$2~weeks to 2~years, with an average duration of $\approx$50~days. The sample under study showed that the duty cycle of transient LMXBs (i.e., the ratio of the time spend in outburst versus that in quiescence) is generally low: on average $\approx$2.5\%, though with a wide spread of $\approx$1--50\%. Figure~\ref{fig:dc_tob_RXTE} shows the distribution of outburst duration (left) and duty cycle (right) for this large sample of LMXBs studied with \rxte.

When comparing the BHs in the sample to the NSs, Yan \& Yu 2015 \cite{Yan2015} found that the BHs peak at higher X-ray luminosity than NSs in absolute numbers. However, when scaled to the mass-dependent Eddington luminosity, the NS  peak at higher ratios (see~Figures~\ref{fig:Lxpeak_RXTE}). Furthermore, it was found in this work that the outbursts of BHs generally decay slower and last longer (88 days on average) than for the NSs (39 days), while the NSs appear to have higher duty cycles. This is shown in Figure~\ref{fig:dc_tob_RXTE}). However, see below for an important caveat about these average numbers.

A number of other, recent works performed similar systematic analyses of large samples of LMXBs, but then focusing specifically on the BH systems. For instance, Dunn et al. 2010 \cite{Dunn2010} studied the behavior of 25 BHs using 13 years of \rxte\ data, and Reynolds \& Miller 2013 \cite{reynolds2013} studied 21 BH LMXBs using 5 years of \swift/XRT data. Furthermore, Tetarenko et al. 2016 \cite{Tetarenko2016a} collected \rxte, \maxi\ and \integral\ data to study 132 outbursts from 57 BH LMXBs (compiled in the WATCHDOG database).\footnote{A complementary BH catalog, called BLACKCAT, has been drawn up by Corral Santana et al. 2016 \cite{CorralSantana16}.}  The average outburst stastics inferred from these BH-targeted studies are broadly consistent with the NS/BH sample analyzed by Yan \& Yu 2015 \cite{Yan2015}.

It is important to note that the above discussed systematic studies do not include i) LMXBs located within a few tens of arminutes (few tens of pc) of the Galactic center, and ii) transient LMXBs with peak luminosities $\lx \lesssim 10^{37}~\dist$~\ergs. This is both because of the limitations in spatial resolution and sensitivity of the instrument that accumulated the large amount of data used in these studies. It is not a priory clear that LMXBs with fainter peak luminosities would behave similar to the brighter systems. In Section~\ref{subsubsec:vfxb} we discuss previous and ongoing efforts to characterize the outburst properties of very faint LMXBs, which have peak outburst luminosities as low as $\lx \simeq 10^{34-36}~\dist$~\ergs, including those located in crowded regions like the Galactic center.

An important word of caution is that the above discussed outburst statistics of LMXBs is typically driven by a small group of sources that have high recurrence rates. For instance, in the work of Yan \& Yu 2015 \cite{Yan2015} the outburst statistics of the NS sample, containing 14 sources in total, is largely driven by the behavior of the 3 most active systems that exhibit outbursts every few ($<$5) years (Aql X-1, 4U 1608--52 and MXB 1730--33). For reference, in the total sample of 36 sources studied by \cite{Yan2015}, $>50$\% exhibited only a single outburst. 

\subsubsection{The role of the orbital period in the long-term X-ray behavior}\label{subsubsec:porb}
The distribution in observed (peak) luminosities among the LMXBs is in part due to the range of orbital periods. Based on theoretical arguments, the peak outburst luminosity is expected to scale with the size of the accretion disk, hence the orbital period of the binary \cite{king1998a,portegieszwart2004,vanHaaften2012}. Although the number of LMXBs with known orbital periods is modest, this theoretical expectation seems to be borne out by observations of samples of NS and BH LMXBs \cite{shahbaz1998,wu2010,vanHaaften2012}. However, a recent study foscussing only on BH LMXBs and applying rigorous statistical tests, did not find a clear correlation between the peak X-ray luminosity and orbital period \cite{Tetarenko2016a}. Other factors that may play a role in setting the outburst peak luminosity, reducing it from the value that could be achieved based on the mass-transfer rate, are mass loss (e.g., via disk winds; \cite{Tetarenko2018}), and local obscuration (e.g., as for BH LMXB V404 Cyg; \cite{koljonen2020}).

Apart from the peak outburst luminosity, the recurrence time of LMXBs may also depend on the orbital period. Mass transfer in LMXBs can be driven by expansion of the donor star, which happens during different evolutionary stages. Systems with long orbital periods ($\gtrsim$12~hr) harbor (sub)giant donor stars that may drive larger mass-accretion rates than the main sequence or evolved donor stars in short-period ($\lesssim$12~hr) LMXBs. This again appears to be largely borne out by observations \cite{Lin2019}.  

Finally, the orbital period of the binary also has impact on whether or not a source is transient or persistent. This distinction depends on the mass-accretion rate and size of the disk; i.e. whether the generated accretion luminosity is sufficient to keep the entire disk ionized hence allowing for continuous accretion  \cite{Coriat2012,intZand2007,Tetarenko2016a}.

\subsection{Short-term X-ray behavior and sub-classes of NS LMXBs}\label{subsec:specstates}
In addition to long-term X-ray variations, both persistent and transient LMXBs vary in X-rays on short timescales: from weeks to days, to second and even sub-second variability. Such short-term variability in X-ray luminosity is often associated with changes in the X-ray spectrum of LMXBs. This gives rise to several distinct X-ray spectral-timing states, which are thought to be related to changes in the accretion morphology \cite{Hasinger1989,Remillard2006,Done2007}

Among the NS LMXBs, several sub-groups with distinct short-term X-ray variability have been defined. For instance, a small group of the brightest NS LMXBs trace out a very distinct ``Z''-shaped pattern when their intensity in different energy bands is plotted in a so-called color-color diagram (CCD). These Z-sources are thought to be accreting around $\lx \simeq \ledd$ (see Section~\ref{subsubsec:vbright}). Other bright NSs, accreting at one to several tens of percent of the Eddington limit, instead trace out an ``C''-shaped pattern in the CCD and are commonly referred to as atoll sources. 


\subsection{Classification based on X-ray luminosity: Two extreme ends}\label{subsec:lum_class}
Wijnands et al. 2006 \cite{Wijnands2006} laid out a classification of LMXBs in terms of their peak X-ray luminosity measured in the 2-10 keV band, making the distinction between bright to very bright ($\lx \sim 10^{37-39}$~\ergs), faint ($\lx \sim 10^{36-37}$~\ergs), and very faint ($\lx \lesssim 10^{36}$~\ergs) systems. This classification is largely driven by the sensitivity limits of historic X-ray instruments and does not necessarily reflect significant physical differences \cite{Wijnands2006}. This is reinforced by the fact that many transient LMXBs display outbursts with different peak luminosity that fall into different luminosity categories \cite{Degenaar2010,Campana2013,Wijnands2013b}. 

Nevertheless, different sub-populations of LMXBs are more likely (or exclusively) found in particular (2-10 keV) luminosity ranges. As already touched upon in previous sections, the X-ray luminosity is expected to scale with the system properties of the LMXB, such as the size of the accretion disk (hence binary orbit) and mode of mass transfer. So there is a physical motivation for a making a distinction based on the peak X-ray luminosity. To illustrate this, we discuss in more detail the two extreme ends of the luminosity classification of LMXBs: very-faint X-ray binaries (VFXBs) and sources that accrete around the Eddington limit. In either of the luminosity classes we find both persistent and transient systems.

\subsubsection{Very-faint X-ray binaries}\label{subsubsec:vfxb}
\begin{figure}
    \centering
    \includegraphics[scale=0.8]{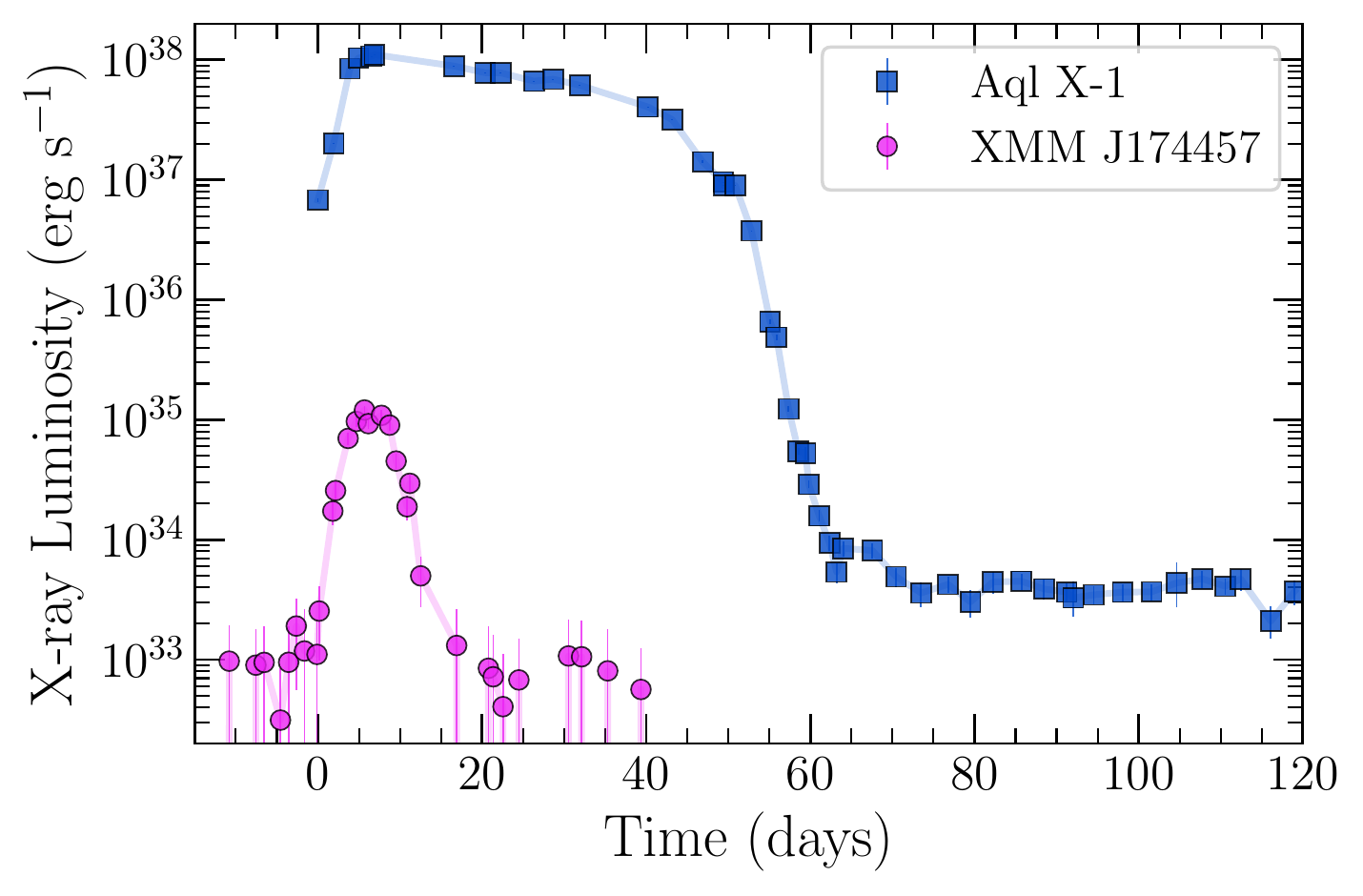}
    \caption{Example lightcurves that illustrate the differences in outburst properties between VFXBs and bright LMXBs. Shown are two frequently two frequently active transient systems that both harbor a neutron star primary: the VFXB XMM~J174457 and bright LMXB Aql X-1. For both sources, Swift/XRT data of outbursts occurring in 2016 were used. It is worth noting that while many transient VFXB outbursts are often short, as in this example, some can last for weeks/months.
}
    \label{fig:vfxt}
\end{figure}

By definition, VFXBs have (peak) accretion luminosities of $\lx \lesssim 10^{36}$~\ergs. Figure~\ref{fig:vfxt} shows an example of how the outburst of a transient VFXB compares to that of a bright LMXB. For sources located at distances of several kpc and beyond, historic X-ray missions and current all-sky X-ray monitors like \swift/BAT and \maxi\ do not have the required sensitivity to detect the accretion emission of VFXBs. This class of LMXBs has therefore long remained hidden. Over the past two decades, however, VFXBs have been found in increasing numbers by various means. At present, a few tens of VFXBs have been identified in the Milky Way. Similar to the bright LMXBs, the VFXBs can be either be classified as (quasi-)persistent or transient based on their long-term X-ray behavior. In Table~\ref{tab:vfxb} we list currently known VFXBs.

A significant number of VFXBs habor NSs and have been discovered through their thermonuclear X-ray bursts, since this makes them shine close to the Eddington limit for a brief moment of time. In particular, the \bepposax\ satellite picked up a number of thermonuclear X-ray bursts without detecting any persistent X-ray counterpart. These objects were hence dubbed ``burst-only sources'' \cite{Cornelisse2002,Cornelisse2002a}. We now know that these concern NSs that accrete at very low X-ray luminosity, either persistently or transiently. Other X-ray instruments such as \integral\ and \swift/BAT have also discovered VFXBs through the detection (and follow-up) of thermonuclear X-ray bursts \cite{Chelovekov2010,Degenaar2010burst,Degenaar2012b}.

The increased sensitivity and improved spatial resolution of X-ray instruments has also allowed for the discovery of a growing number of VFXBs through X-ray surveys, primarily of the Galactic center and bulge, with missions such as {\it ASCA}, \chandra\ and \xmm\ \cite{Sakano2005,Muno2005,Wijnands2006,Jonker2011,Degenaar2012a}. However, a detailed characterization of the outburst properties of VFXBs, such as their outburst duration and recurrence time, requires monitoring surveys in regular (at least weekly) visits with sensitive X-ray instruments to be performed \cite{Degenaar2015GC,Carbone2019,Bahramian2021}. Such programs have been carried out mostly with \rxte/PCA and \swift/XRT.

The \rxte/PCA bulge monitoring project ran between 1999 and 2011 \cite{Swank2001} and pushed to a factor of a few lower X-ray luminositites than typical all-sky X-ray monitors ($\lx$ of a few times $10^{35}$~\ergs). This allowed to study the behavior of the brightest VFXBs and provided a first picture of their outburst durations and recurrence times. In more recent years, this has been complemented by monitoring campaigns that exploit the flexibility of the \swift\ mission and push down a further order of magnitude fainter X-ray luminosity \cite{Degenaar2015GC,Bahramian2021}.


Statistics on the outburst properties of VFXBs has been steadily accumulated through \swift's Galactic Center monitoring program, which has provided nearly-daily X-ray monitoring of about a dozen transient VFXBs since 2006 \cite{kennea2006,Degenaar2015GC}. Initial studies using the first few years of monitoring data showed that the outbursts of the transient VFXBs are often short ($\lesssim$1~month), but that the outburst recurrence times are not very different from that of bright LMXBs: typical duty cycles are $\simeq$1--30\% \cite{Degenaar2009,Degenaar2010}. As the \swift\ Galactic center monitoring program is reaching a $\simeq$15~yr baseline, it is starting to provide similar outburst statistics on transient VFXBs as has been available for bright transient LMXBs. 

Apart from providing valuable statistics on the X-ray outburst behavior, determining the system parameters of VFXBs (i.e., donor types and orbital periods) in the Galactic center is  largely hampered by the very high extinction and crowding. Based on what has been learned from the outburst behavior of VFXBs in the Galactic center, a dedicated monitoring program was therefore designed to effectively find such systems with \swift/XRT in less obscured regions of the Galactic bulge \cite{Shaw2020,Bahramian2021}. This \swift\ bulge survey is ongoing and are expected to provide more insight into the (distribution of) system parameters of the faintest accreting LMXBs (see Section~\ref{subsec:galbulge}).

Despite the observational challenges, system parameters are known for a number of VFXBs.  
Nearly half of the VFXBs listed in Table~\ref{tab:vfxb} have NS primaries, as evidenced by the detection of thermonuclear X-ray bursts or coherent X-ray pulsations. In absence of such features it is often difficult to probe the nature of the compact object. However, dynamical mass estimates reveal BHs in Swift J1357.2--0933 \cite{CorralSantana2013} and XTE J1118+480 \cite{McClintock2001a,wagner2001}. Furthermore, the X-ray spectral properties of XTE J1728--295 point to a BH accretor \cite{Sidoli2011,Stoop2021}, and the bright radio emission of the Galactic center transient CXOGC J174540.0--290031 seem to argue in favor of a BH as well \cite{Muno2005a,porquet2005}. The latter is, however, a high-inclination system that while appearing as a VFXB might be intrinsically bright.

At present, the compact object has not yet been classified for nearly half of VFXBs. The challenge arises from the fact that thermonuclear X-ray bursts are rare at low accretion rates (i.e., it takes a long time to accumulate sufficient fuel to ignite a burst; \cite{Degenaar2010burst,degenaar2011burst}), and X-ray pulsations are difficult to detect at low X-ray fluxes \cite{patruno2010b,Strohmayer2017,vandeneijnden2018,bult2019,bult2021}. Furthermore, the donor stars are often too faint or too highly absorbed for optical mass measurements in quiescence \cite{Shaw2017}. Similarly, indirect methods such as studying their rapid incoherent X-ray variability is hampered by their faint emission \cite{armaspadilla2014}. Therefore, other methods have been explored to classify VFXBs.

When the outburst of a transient VFXBs is densely monitored (e.g., using \swift) or high-quality X-ray spectra are obtained (e.g., with \xmm), a tentative identification of the compact object may be made based on the presence of a soft spectral component \cite{ArmasPadilla2011} (albeit see \cite{ArmasPadilla2013} and \cite{ArmasPadilla2013a}), or the X-ray spectral evolution along the decay of an accretion outburst \cite{Wijnands2015}.  At least one transient VFXB identified as NS based on this method (IGR J17494--3030; \cite{ArmasPadilla2013b,Wijnands2015}) was later found to be an AMXP \cite{Ng2021}. Spectral shape and its evolution can thus be a diagnostic to identify the compact object in VFXBs (though see \cite{parikh2017} for NSs with unusually hard spectra that do not confirm the general trends). 

For several VFXBs the detection of coherent (millisecond) X-ray pulsations has allowed 
to measure orbital period through pulsar timing
 \cite{Sanna2018a,Strohmayer2018}, and for a handful of others such information is available from X-ray or multi-wavelength studies \cite{Bahramian2014,Shaw2020}. 
This suggests that VFXBs are an inhomogenous class in terms of donor stars and mass-transfer mode. The population of VFXBs includes systems with evolved dwarf donors in short orbits, main sequence and giant donors in wider orbits, Roche-lobe overflow and wind-fed systems, as well as various sub-classes of LMXBs such as UCXBs and tMSRPs.\footnote{We note that other classes of compact objects such as HMXBs, magnetars and bright accreting white dwarfs (cataclysmic variables) can show up as X-ray transients with similar luminosities as that of transient VFXBs. The X-ray spectral-timing properties of multi-wavelength properties of such objects would be different from LMXBs, which are the focus here.}  

So far, many of the known VFXBs have (X-ray) characteristics consistent with Roche-lobe overflow LMXBs, so this appears to the dominant sub-class. Since the accretion luminosity of LMXBs scales (among other factors) with the orbital period (see Section~\ref{subsec:porb}), it is not surprising that several VFXBs have been found to have short orbital periods. For instance, the few BH systems among the VFXBs all have  orbital periods shorter than BHs in bright LMXBs (see Section~\ref{subsec:porb}). Furthermore, several VFXBs have been identified as (candidate) UCXBs (see Table~\ref{tab:vfxb}). Binary evolution considerations predict that VFXBs should present short-period systems with evolved donor stars \cite{King2006}

There are also a few (transient or persistent) VFXBs that are SyXBs, e.g., accreting from the wind of a giant companion \cite{Shaw2020,Bahramian2021}. This is also not unexpected based on binary evolution arguments \cite{Maccarone2013}. Finally, we note that a number of VFXBs may be intrinsically bright LMXBs that only appear sub-luminous because we are viewing the system at high inclination. One example is the eclipsing Galactic center transient VFXB CXOGC J174540.0--290031 \cite{Muno2005a,porquet2005}. However, only a small fraction of the full VFXB population is expected to appear faint due to such inclination effects \cite{Wijnands2006}.

Summarizing, separating VFXBs as a class is mostly phenomenological, but there might be actual physical differences with respect to brighter systems; e.g., they might be less X-ray luminous because their accretion disks are small overall, or because their inner accretion disk is evaporated into a radiatively-inefficient flow or pushed out by the NS magnetic field. Moreover, at the low accretion rates associated with VFXBs, we can probe different physics than with bright LMXBs in terms of accretion and outflows \cite{degenaar2017,vandeneijnden2021}, NS physics \cite{wijnands2008}, thermonuclear burning \cite{Cooper2007,Peng2007}, and binary evolution \cite{maccarone2015}.

\begin{table}
\centering
\begin{tabular}{lcccc}
\hline
\hline
System & T/P/QP  & NS/BH & Comments & Reference \\
\hline
\multicolumn{5}{c}{{\bf Persistent and quasi-persistent VFXBs}}\\
IGR J17254--3257 & P & NS & burster &  \cite{chenevez2007,Ratti2010} \\
1RXS J171824.2--402934 & P & NS & cUCXB, burster & \cite{intZand2005a,intZand2009}\\ 
1RXH J173523.7--354013 & P & NS & burster, H-rich donor &  \cite{Degenaar2010burst}\\
M15 X-3 & P & cNS & main-sequence donor & \cite{Heinke2009a,Arnason2015} \\
IGR J17062--6143  & QP & NS & UCXB, AMXP, buster & \cite{Strohmayer2018,Hernandez2019} \\ 
AX J1754.2--2754  & QP & NS & cUCXB, burster &  \cite{Chelovekov2007,Shaw2017a} \\ 
XMMU J174716.1--281048 & QP & NS & cUCXB, burster & \cite{DelSanto2007,Kaur2017}\\
SAX J1806.5--2215 & QP & NS & burster & \cite{Cornelisse2002,Shaw2017a} \\
IGR J17597--2201 & QP & NS & buster, XTE J1759--220 &   \cite{Brandt2007,Fortin2018} \\
AX J1538.3--5541  & QP & ? &  LMXB-like spectrum  & \cite{Sugizaki2001,degenaar2012asca}\\ 
XTE J1744--230 & QP & ? & RXTE bulge scan &  \cite{Swank2001} \\ 
\hline
\multicolumn{5}{c}{{\bf Transient VFXBs}}\\
IGR~J17379--3747$^a$ & T & NS  & burster, AMXP & \cite{Chelovekov2010,Sanna2018b} \\ 
IGR~J17591--2342 & T & NS & burster, AMXP & \cite{Sanna2018a,Kuiper2020} \\
Swift J185003.2--005627 & T & NS & burster & \cite{Degenaar2012b} \\
Swift J1734.5--3027 & T & NS & burster & \cite{Bozzo2015a} \\
MAXI J1807+132 & T & NS & burster &  \cite{Shidatsu2017,Jimenezibarra2019} \\
MAXI J1957+032 & T & cNS &  &  \cite{MataSanchez2017,Beri2019} \\
IGR J17494--3030 & T & cNS &  &  \cite{ArmasPadilla2013b,Ng2021} \\ 
XTE J1719--29 & T & cNS &  & \cite{ArmasPadilla2011} \\ 
XTE J1118+480 & T & BH  & $P_{\mathrm{orb}}=$~4.1~hr & \cite{wagner2001,McClintock2001a} \\
Swift J1357.2--0933 & T & BH  & $P_{\mathrm{orb}}=$~2.6~hr & \citenum{ArmasPadilla2013b,CorralSantana2013,MataSanchez2015} \\
XTE J1728--295 & T & cBH & IGR J17285--2922 &  \cite{Swank2001,Sidoli2011,Stoop2021} \\
IGR J18175--1530  & T & ? & XTE J1817--155 & \cite{Markwardt2007b} \\ 
XMMSL1 J171900.4--353217 & T & ? & XTE J1719--356 & \cite{Read2010a,Markwardt2010b} \\
XTE J1734--234 & T & ? &  &  \cite{Swank2001} \\ 
WGA J1715.3--2635 & T & ? &  & \cite{Swank2001} \\ 
XTE J1637--498 & T & ? &  & \cite{Markwardt2008} \\ 
Swift J175233.9–-290952 & T & ? &  & \cite{Shaw2020,Bahramian2021}\\
Swift J174038.1--273712 & T & ? &  & \cite{Bahramian2021a,RiveraSandoval2021}\\
SRGt J071522.1--191609 & T & ? & \cite{Gokus2020,vandenEijnden2020_vfxt,Bahramian2020a} \\
IGR~J17445$-$2747   & T & NS & SyXB & \citep{Mereminskiy2017,Shaw2020,Bahramian2021}\\
XMMU~J174445.5$-$295044 & T & ? & SyXB & \cite{Heinke2009b,Bahramian2014a}\\
\hline
\multicolumn{5}{c}{{\bf VFXBs covered by Swift's Galactic Center monitoring campaign}}\\
XMM J174457--2850.3 & T & NS & burster & \cite{Sakano2005}    \\
Swift J174622.1--290634  & T & ? &  &  \cite{Degenaar2009}   \\ 
Swift J174553.7--290347   & T & ? & CXOGC J174553.8--290346 &  \cite{Degenaar2009}  \\ 
CXOGC J174540.0--290005   & T & ? &  &  \cite{Muno2005,Degenaar2009,Koch2014,Heinke2015}   \\ 
CXOGC J174538.0--290022  & T & ? &  & \cite{Muno2005,Degenaar2015GC}   \\ 
CXOGC J174535.5--290124  & T & ? &  &   \cite{Muno2005,Degenaar2009,Degenaar2010}  \\ %
CXOGC J174541.0--290014  & T & ? &  &   \cite{Muno2005}   \\  
CXOGC J174540.0--290031  & T & cBH & $P_{\mathrm{orb}}=$7.9~hr, eclipser, obscured? & \citenum{Muno2005,Muno2005a,porquet2005}   \\  
CXOGCJ174554.3--285454 & T & ? & XMMU J174554.4--285456 &    \cite{Muno2005}    \\
XMM J174544--2913.0 & T & ? &  &   \cite{Sakano2005}   \\ 
Swift J174535.5--285921  & T & ? &  &   \cite{Degenaar2015GC}   \\ 
\hline
\end{tabular}
\caption{List of VFXBs in our Galaxy. The letters P, QP and T refer to persistent, quasi-persistent and transient, respectively. cBH and cNS refer to candidate BHs and candidate NSs, respectively. The indication ``burster'' implies that the source displays thermonuclear X-ray bursts (hence must harbor a NS). $^a$ = This is the same source as IGR~J17364--2711, IGR~J17380--3749 and XTE~J1737--376.}
\label{tab:vfxb}
\end{table}

\subsubsection{Accretion around the Eddington luminosity in LMXBs}\label{subsubsec:vbright}
At the other extreme end of VFXBs in terms of X-ray output, there are a modest number of LMXBs that accrete around (slightly below and above) the Eddington luminosity. Table~\ref{tab:Edd} lists some of the basic characteristics of LMXBs that are known to accrete at these high rates. Most of these harbor (confirmed or expected) NS primaries; there are only two BH-LMXBs listed. These are the well-known quasi-persistent BH GRS 1915+105 (see Figure~\ref{fig:lcs} bottom), and the transient system V404 Cyg (see Figure~\ref{fig:binsim} middle left). The latter is not observed to be particularly luminous, but is thought to be heavily obscured and intrinsically accreting around the Eddington limit during (some of) its outbursts \cite{Tetarenko2016a, Motta2017, koljonen2020}. 

The group of Eddington-accreting LMXBs includes the 6 persistent Z-sources (see also Section~\ref{subsec:specstates}), which have been known and extensively studied since the dawn of X-ray astronomy \cite{Hasinger1989}. These sources are expected to harbor NS primaries, although only GX 17+1 and Cyg X-2 show conclusive evidence for this by the detection of thermonuclear bursts \cite{Kahn1984}. Identifying the donor stars in the Z-sources is not straightforward, because they are persistently accreting and their accretion disks and jets are so bright that these contribute significantly to the optical/infrared emission \citep{Bandyopadhyay2003}. A schematic impression from one of the Z-sources, Sco X-1, is shown in Figure~\ref{fig:binsim} (middle right). 
Furthermore, Cir X-1 and GX 13+1 are two persistent sources that exhibit variations in their X-ray luminosity but during some epochs behave similar to the Z-sources \cite{Oosterbroek1995}. Both Cir X-1 and GX 13+1 are known to harbor a NS through the detection of thermonuclear bursts \cite{Matsuba1995,Linares2010}. In case of Cir X-1, it is debated whether it is a LMXB or a HMXB system \cite{Johnston2016}.

There are also 4 transient LMXBs, confirmed to harbor NSs through the detection of thermonuclear bursts or quiescent thermal emission, that showed similar behavior as the Z-sources during the peak of their outbursts (see Table~\ref{tab:Edd}). These are, therefore, thought to have reach accretion rates around the Eddington limit (despite their distances being poorly constrained in some cases), and are referred to as transient Z-sources \cite{Wijnands1999a,altamirano2010_z,Homan2010,Homan2014}. Furthermore, GRO J1744--28 is a transient LMXB that harbors a slowly pulsating neutron star \cite{Finger1996} accretes around the Eddington limit at the peak of its outbursts \cite{Monkkonen2019}.

Finally, there are a handful of other LMXBs that, like V404 Cyg, are not observed to be particularly luminous but are suspected to be intrinsically accreting around the Eddington limit. Evidence pointing in this direction comes e.g., from their orbital period evolution \cite{baknielsen2017}, or their strong jets and disk winds \cite{panurach2021}. For these sources it appears that the most luminous inner part of the binary is obscured, e.g., due to a high inclination or because of strong disk winds.

As can be seen in Table~\ref{tab:Edd}, several of the high-accretion rate LMXBs have long orbital periods and evolved donor stars. Indeed, their orbital periods are typically too long for a main sequence star to fill its Roche lobe and drive the observed high mass accretion rates. As discussed in Section~\ref{subsubsec:porb}, finding that  bright LMXBs have relatively long orbital periods is theoretically not surprising. Indeed, it was already suggested early on that the mass donors in Z-sources are evolved stars that are ascending on the giant branch and that high mass-transfer rates could be driven by the associated expansion of the donor star \cite{Taam1983,Webbink1983,Hasinger1989}. However, at least one Z-source likely cannot harbor an evolved donor; the infrared counterpart of GX 17+1 is far too faint to allow a giant donor unless the source distance is wrong \cite{jonker2000,Callanan2002}. It should also be noted that large orbital period and evolved companion star do not necessarily imply that mass-accretion can be driven up to the Eddington limit. For instance, 4U 1624--49 likely hosts an evolved star and has an orbital period of $\approx$21~hr \cite{Jones1989}, but is a luminous atoll instead of a Z-source.  

We note that similarly high accretion rates as encountered in these LMXBs are also reached in some HMXBs. Well-known examples are the (c)BH systems SS443, V4641 Sgr, and Cyg X-3 \cite{koljonen2020}. Furthermore, (super-)Eddington accretion rates are a defining property of the population of ultra-luminous X-ray sources (ULXs). These are extra-Galactic BHs or NSs that appear to be accreting at super-Eddington rates and of which there are many hundreds known to date \cite{Walton2022}. However, a large fraction of ULXs may be HMXBs instead of LMXBs (see \cite{Kuranov2021} and references therein). 

\begin{table}
\centering
\begin{tabular}{lcccccc}
\hline
\hline
System & T/P/QP  & NS/BH & $P_{\mathrm{orb}}$ (d) & Donor & Comments & Reference \\
\hline
Cyg X-2 & P & NS & 9.8 & A giant  & Z-source & \cite{Cowley1979,Casares1998} \\
GX 349+2 & P & NS & 0.9  & sub-giant?  & Z-source & \cite{Wachter1996,Bandyopadhyay1999} \\
Sco X-1 & P & NS & 0.8 &  sub-giant (K or later) & Z-source & \cite{lasala1985,MataSanchez2015b} \\
GX 5-1 & P & NS & ? & early-type giant?  & Z-source & \cite{jonker2000,Bandyopadhyay2003} \\
GX 17+1 & P & NS & ? & ?  & Z-source & \cite{jonker2000,Callanan2002} \\
GX 340+0 & P & NS & ? & ?  & Z-source & \cite{vanderKlis2006} \\
\hline
IGR J17480--2446  & T & NS & 0.9 &  sub-giant & transient Z-source, AMXP   & \cite{altamirano2010_z,Patruno2012b} \\
XTE J1701--462 & T & NS & ?  & ? & transient Z-source & \cite{Homan2010} \\
MAXI J0556--332 & T & NS & ?  & ? & transient Z-source & \cite{Homan2014} \\
2S 1803--245 & T & NS &  ? & ? & transient Z-source & \cite{Wijnands1999a} \\
\hline
GX 13+1 & P &   NS & 24.5 & K giant  & hybrid atoll/Z-source & \cite{Garcia1992} \\ 
Cir X-1 & P &  NS & 16.7  & debated  & often in low-flux state & \cite{Shirey1999,Johnston2016} \\
GRO J1744--28 & T &  NS & 11.8  & G/K giant  & slow pulsar & \cite{Finger1996,Gosling2007}\\
\hline
GRS 1915+105 & QP & BH & 33.5  &  K giant & sometimes obscured & \cite{Tetarenko2016a} \\
V404 Cyg & T & BH & 6.5 & K subgiant  & obscured & \cite{koljonen2020} \\
2A 1822--371 & P & NS & 0.23 & main sequence  & obscured, slow pulsar & \cite{Jonker2003b,baknielsen2017}  \\
Swift J1858.6--0814 & T & NS & 0.9 & ?  & obscured & \cite{Buisson2020,Buisson2021} \\
AC 211 (X2127+119) & P & NS? & 0.7 & ?  & obscured & \cite{vanZyl2004,panurach2021} \\
\hline
\end{tabular}
\caption{List of Galactic LMXBs accreting around the Eddington limit. The letters P, QP and T refer to persistent, quasi-persistent and transient, respectively. We note that in case of Cir X-1 it is debated whether the system is a LMXB or a HMXB system \cite{Johnston2016}.}
\label{tab:Edd}
\end{table}


\section{Distribution and demographics of LMXBs in the Galaxy}\label{sec:distr_demo}
\begin{figure}
    \centering
    \includegraphics[scale=0.5]{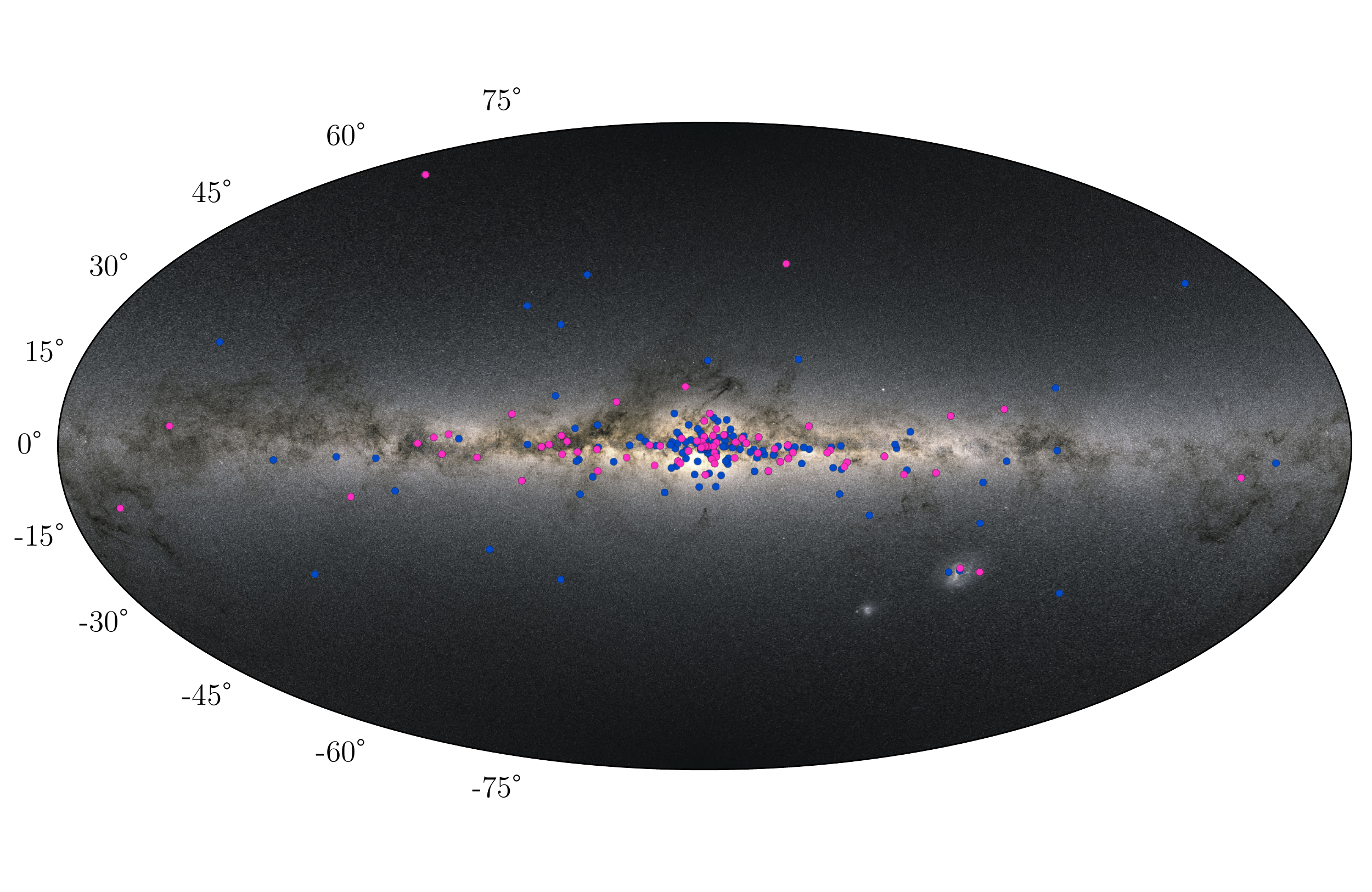}
    \caption{Distribution of a subset of known LMXBs in the sky overlaid on the Gaia all-sky image. Pink circles represent BH-LMXBs from \citep{Tetarenko2016a}, and blue circles are all other LMXBs as cataloged by 
    \citep{Liu2007}. Given discovery of new LMXBs over the past decade and new revelations on classification of some previously known LMXBs, this plot is not representing the entire currently known sample of LMXBs and it serves only as an example demonstrating the Galactic distribtuion of LMXBs. Background image from the \textit{Gaia} mission (A. Moitinho; ESA/Gaia/DPAC). Figure made using \textsc{mw-plot} (\url{https://pypi.org/project/mw-plot/}).}
    \label{fig:galplot}
\end{figure}

LMXBs are found all around us in the Galaxy (Figure~\ref{fig:galplot}), however their formation channel, population density, and demographic varies depending on the environment. Formation of the compact object in X-ray binaries inflicts a kick on the binary. This kick adds a peculiar motion to the X-ray binary's Galactocentric orbit and can displace it substantially from where the binary was formed. Despite this, given the large mass and short lifespan of the companion, HMXBs are not generally expected to be displaced substantially from their original birthplace. Observations indicating HMXBs trace star formation and young star associations support these expectations \citep{Grimm2003, Bodaghee2012}. In contrast, the low mass and long lifespan of companions in LMXBs allows them to achieve significant displacement  from their birthplace in present day observations  \citep[e.g.,][]{Repetto2012} or be found in regions associated with old star populations \citep[e.g., Galactic center, Galactic bulge, Globular clusters, e.g.,][]{Arnason2021}.

Observed spatial distribution of LMXBs also shows a strong concentration of these systems in the central 2 kpc of the Galaxy, with the overal population showing a larger scale height compared to HMXBs \citep[410~pc for LMXBs, compared to 150~pc for HMXBs][]{Grimm2002}. Consequently, \cite{Grimm2002} also note that around $\sim50\%$ of LMXBs are in the Galactic disk, compared to $\sim25\%$ in the Galactic bulge.

Over the past few decades monitoring and survey programs with various X-ray observatories have provided a more complete picture of distribution of LMXBs in our Galaxy. These include All-sky monitoring programs \citep[e.g., with \rosat, \swift/BAT, \maxi, \integral,][]{Voges1999, Baumgartner2013, Kawamuro2018, Hori2018, Krivonos2013}, and survey programs targeting the Galactic center \citep{Muno2006a, Heard2013, Degenaar2015GC, Sazonov2020}, Galactic bulge \citep{Jonker2014, Kuulkers2007, Grebenev2015, Bahramian2021}, Galactic plane \citep{Giacconi1971, Grindlay2003, Motch2010}, and globular clusters \citep{Clark1975, Verbunt2001, Grindlay2001a, Pooley2003}. These surveys have substantially pushed our understanding of LMXB population and distribution in the Galaxy. However, it is important to note that our current view is still heavily impacted by selection effects caused by telescope sensitivity and survey cadence limitations, interstellar absorption \citep{Jonker2021}, among other effects such as uneven coverage. Current and future survey efforts (e.g., \textit{e-ROSITA}) will improve on some of these limits. In the following sections, we briefly discuss a subset of surveys and monitoring programs targeting different parts of the Galaxy in the context of LMXBs.

\subsection{Galactic center and bulge}\label{subsec:galbulge}
The Galactic center has been heavily observed by various X-ray observatories including \chandra, \xmm, \swift, \integral. Using \chandra\ \citep{Wang2002,Muno2006a} reported and cataloged hundreds of X-ray sources in the central part of the Galaxy, identifying numerous X-ray sources associated with old stellar populations, thus classifying a large fraction of them as old cataclysmic variables, and a smaller group ($\sim10$) as likely LMXBs. \xrt\ has been observing the Galactic center on a $\sim$weekly basis since 2006. These observations have led to identification of multiple transient LMXBs and outbursts within $25'$ of the Galactic center \cite{Degenaar2015GC}. A large fraction of these transients were also identified as VFXBs, showing outbursts with peak L$_X < 10^{36}$ \ergs\ (see Section~\ref{subsubsec:vfxb}). 

Identification of dozens of faint and transient likely LMXBs (and hundreds of other unclassified X-ray sources) in the Galactic center motivated theoretical exploration of the origin and nature of these energetic systems in the region. Population synthesis work by \cite{Liu2006} suggested that NS-LMXBs are likely to dominate the population of X-ray sources in the Galactic center. However, later work by \cite{Ruiter2006} suggested that a majority of these sources may be intermediate polars (a subclass of cataclysmic variables) and that the population of LMXBs is likely to be comparatively small ($\sim$dozens). This was found to be consistent with further observations of the Galactic center by \chandra\ \cite{Zhu2018} suggesting number of quiescent LMXBs in the Galactic center cannot be substantially larger than $\sim 150$.

\cite{Muno2005} noted an overabundance of transients (particularly transient VFXBs) in the central parsec of the Galaxy - suggesting these systems could be LMXBs that have formed dynamically (similar to globular clusters; see Sec~\ref{subsec:globclust}), or alternatively HMXBs associated with the young star population in the central region. However, this overabundance has not been observed in the \xrt\ monitoring of the Galactic center out to 25~pc (running since 2006; \cite{Degenaar2015GC}).

Extending X-ray surveys to the Galactic bulge, the \chandra\ Galactic Bulge Survey identified $\sim170$ candidate cataclysmic variables and quiescent LMXBs \citep{Jonker2011, Jonker2014, Wevers2016, Wevers2017}. \xrt\ monitoring of the Galactic bulge with, led to identification of $\sim$7 new VXFBs, most of which appear to be symbiotic in nature \citep{Shaw2020, Bahramian2021}; leading them to conclude that a large fraction of VFXBs are perhaps SyXBs. At higher luminosities, \integral\ surveys of the Galactic bulge have identified dozens of bright LMXB candidates \citep{Kuulkers2007,Grebenev2015}. Exploring characteristics of the X-ray luminosity function of persistent versus transient LMXBs detected by \integral\ \cite{Revnivtsev2008} concluded that the distribution of transient LMXBs is comparatively more focused around the Galactic center. 

\subsection{Galactic plane and outer parts}\label{subsec:galplane}
Survey and monitoring coverage of the Galactic plane and halo in the X-rays is substantially shallower and less complete than the Galactic center and bulge. Nevertheless, surveys such as the \xmm\ or \swift\ Galactic Plane surveys \citep{Motch2010, Gorgone2019} have discovered a handful of LMXBs and LMXB candidates. However, the overwhelming majority of LMXBs identified in the Galactic plane and halo are discovered while exhibiting bright outbursts detected by all-sky monitors such as \maxi\ or \swift/BAT.

Most LMXBs (outside globular clusters) are expected to have formed in the Galactic plane (e.g., following the mass distribution of low-mass stars in the Galaxy \cite{Gilfanov2004}). However, a small number of LMXBs are found at high Galactic latitudes \cite[e.g., see][]{Jonker2004b, shaw2013, Arnason2021}. While it is possible that a non-zero fraction of LMXBs have formed in the halo, a more likely explanation is receiving high natal kicks when the BH or NS is formed through supernova. Given the relatively old age of LMXBs, a high binary natal kick can displace the LMXB significantly from where the compact object in the binary was originally formed\footnote{Assuming the binary is not disrupted as a result of the supernova kick.} \cite{Repetto2012}. 

While both BH-LMXBs and NS-LMXBs have been found with high Galactic scale heights, indicating evidence for large kicks in both classes \citep{repetto2015}, NS-LMXBs are found to reach higher scale heights compared to BH-LMXBs with similar binary natal kicks \citep{repetto2017}. Furthermore, comparing observation-based distribution of binary natal kicks of BH-LMXBs and NS-LMXBS indicates that the kick velocities are smaller for BH-LMXBs compared to NS-LMXBs \citep{Atri2019}.

\subsection{Globular clusters}\label{subsec:globclust}
Observationally, an overabundance of X-ray binaries in Milky Way globular clusters was noticed soon after the emergence of X-ray astronomy \citep{Clark1975}. Globular clusters contain $\sim10^{-4}$ fraction of Galactic mass, while containing $\sim10$--$20\%$ of the total population of Galactic LMXBs.

In the Galactic field, LMXBs are thought to have predominantly formed through ``primordial'' binary evolution \citep[e.g.,][]{Paczynski76, Tauris2003} (however, challenges remain in explaining formation of LMXBs with heavy compact objects - e.g., see \citep{Podsiadlowski2003}). In contrast, the high density of globular clusters (which can reach up to $10^6$ times the Solar neighborhood), leads to frequent encounters between members of the cluster. These encounters are expected to open multiple channels for formation of LMXBs in globular clusters, including collision of red giants with a compact object \citep{Sutantyo1975}, tidal capture of a compact object by a companion star \citep{Fabian1975}, or exchange encounters during which a compact object replaces the lower mass member of an existing binary \citep{Hills1976}. 

Observational evidence supporting this link was first discussed in \citep{Verbunt1987}. Detailed follow up studies with the \chandra\ observatory allowed high resolution imaging of faint X-ray sources and provided clear evidence of this link in the Milky Way \citep{Pooley2003, Heinke2003c} and other galaxies \citep{Sarazin2003, Kundu2007, Kundu2007a}. Focused studies of individual globular clusters over the past two decades have led to identification and classification of dozens of X-ray binaries \citep[e.g.,][]{Grindlay2001, Heinke2006a, Servillat2008, Maxwell2012}, and X-ray emission from radio millisecond pulsars \citep{Bogdan2010, Bhattacharya2017}. Furthermore, a non-negligible fraction of transient and persistent LMXBs (with peak L$_X >10^{35}$\ergs) discovered in the Milky Way have been localized to globular clusters \citep[see Table~\ref{tab:gcLMXBs},][]{Bahramian2014}. A large fraction of these systems have been determined to be NS-LMXBs based on detection of type I X-ray bursts or of X-ray pulsations. 

There are also numerous quiescent LMXBs identified in globular clusters which have not exhibited bright outbursts yet. These systems are identified through X-ray spectroscopy, generally showing a spectrum well described by NS atmospheric models \citep[e.g.,][]{Rutledge2002, Gendre2003a, Becker2003, Heinke2006, Guillot2009, Walsh2015, Bahramian2015}. 

Quiescent and transient NS-LMXBs in globular clusters provide unique opportunities for study of accretion and NS equation of state. These are largely thanks to the independent constraints on distance that is achieved for globular clusters (e.g., see \citep{Harris1996} and references therein for distance measurements) and in some cases existence of pre-outburst observations of the LMXB as the globular cluster was targeted for other studies. Observations of quiescent NS-LMXBs have allowed detailed study of the equation of state in NSs based on constraints on mass and radius obtained via X-ray spectral modeling of the NS atmosphere, which is the dominant emitting region in the X-rays with little or no contamination from the accretion flow in quiescence \citep{Guillot2013, Heinke2014, Steiner2018}. Detailed study of transient systems during and after their outburst decay has provided insights about thermal processes in the crust of NSs, for example showing that internal heating of NSs occurs significantly closer to the surface than previously thought \citep[e.g.,][]{Degenaar2011b, Degenaar2012, wijnands2017, Vats2018}. Multi-wavelength observations of transient LMXBs in globular clusters during their outbursts has led to the first confirmation of a transitional millisecond pulsar \citep{Papitto2013}, and strong constraints on the disk-jet coupling in NS-LMXBs \citep{Tetarenko2016}.

In contrast with the Galactic field, where there are dozens of strong BH-LMXB candidates with $\sim$20 dynamically confirmed \citep{Tetarenko2016a, CorralSantana16}, there are currently no dynamically confirmed BH-LMXBs identified in globular clusters (Galactic or otherwise), and almost all transient LMXBs in globular clusters so far have been identified to be NS-LMXBs (Table~\ref{tab:gcLMXBs}, \citep{Verbunt2006}). Historically, it was thought that while globular clusters produce thousands of BHs through their evolution, a large fraction of these BHs decouple from the rest of the cluster in form of a subcluster, segregate towards the cluster core and eventually get ejected from the cluster \citep[e.g.][]{Sigurdsson1993}. However, detection of strong BH-LMXB candidates in Galactic \citep{Strader2012, Chomiuk2013} and extra-galactic clusters \citep{Maccarone2007, Dage2020}, along with identification of detached dynamically confirmed BHs \citep{Giesers2018, Giesers2019}, changed these assertions. 

\begin{table*}
\begin{center}
\begin{tabular}{lccccccc}
\hline
\hline
LMXB	 		& globular cluster & State & P$_{orb}$ & nature & notes & accretor & references\\
\hline
\multicolumn{8}{c}{\textbf{Bright LMXBs (peak L$X\geq 10^{35}$ \ergs)}}\\
4U 1820-30 		& NGC 6624	& P	& 11 min 		& U & X  & NS & \citep{Stella1987}\\
4U 0513-40 		& NGC 1851	& P	& 17 min		& U & UV & NS & \citep{Zurek2009}\\
X1850-087  		& NGC 6712	& P	& 20.6$^a$ min  & U & UV & NS & \citep{Homer1996}\\
M15 X-2 	    & M 15 	 	& P	& 22.6 min	    & U & UV & NS & \citep{Dieball2005}\\
MAXI J0911-655 	& NGC 2808 	& T & 44.3 min   	& U & XP & NS & \citep{Sanna2017}\\
IGR J16597-3704 & NGC 6256 	& T & 46.0 min      & U & XP & NS & \citep{Sanna2018}\\
NGC 6440 X-2 	& NGC 6440	& T	& 57.3 min	    & U & XP & NS & \citep{Altamirano2010}\\ 
XB 1832-330 	& NGC 6652 	& P	& 2.1 hrs 		& N & O  & NS & \citep{Engel2012}\\ 
4U 1746-37 		& NGC 6441 	& P	& 5.16 hrs 	    & N & X  & NS & \citep{BalucinskaChurch2004}\\
SAX J1748.9-2021& NGC 6440 	& T	& 8.7 hrs 		& N & XP & NS & \citep{Altamirano2008}\\ 
IGR J18245-2452 & M28 	 	& T	& 11.0 hrs		& N & XP & NS & \citep{Papitto2013}\\
GRS 1747-312 	& Terzan 6 	& T	& 12.36 hrs 	& N & X  & NS & \citep{intZand2003}\\
AC 211 			& M 15  	& P	& 17.1 hrs 	    & N & UV & NS? & \citep{Ilovaisky1993}\\
IGR J17480-2446 & Terzan 5 	& T	& 21.27 hrs	    & N & XP & NS & \citep{Papitto2011}\\
Rapid Burster	& Liller 1	& T	& ?			    & N & B  & NS & \citep{Galloway2008} \\
EXO 1745-248	& Terzan 5 	& T	& ?			    & N & B  & NS & \citep{Galloway2008} \\
Terzan 5 X-3	& Terzan 5 	& T	& ?			    & N & B  & NS & \citep{Bahramian2014}\\
XB 1732-304     & Terzan 1  & T & ?			    & ? & ?  & NS & \citep{Guainazzi1999}\\
4U 1722-30      & Terzan 2  & P & ?			    & U?& B  & NS & \citep{intZand2007}\\
IGR J17361-4441 & NGC 6388  & T & ?  	        & ? & ?  & NS? & \citep{Bozzo2011}\\
CXOU J173324.6-332321 & Liller 1 & T & ?        & ? & ?  & NS? & \citep{Homan2018} \\
MAXI J1848-015  & GLIMPSE C01 & T & ?           & ? & ?  & ? & \citep{Kennea2021}\\
\hline
\multicolumn{8}{c}{\textbf{Faint LMXBs (peak L$X\leq 10^{35}$ \ergs)}}\\
47 Tuc X9 	    & 47 Tuc    & V &  28.2 min     & U &	X & BH? & \citep{MillerJones2015, Bahramian2017}\\
M15 X-3 	    & M 15 	    & T &  $\sim240$ min& ? &	AC& NS? & \citep{Heinke2009a, Arnason2015}\\
NGC 6652 B 	    & NGC 6652 	& V &  ? 	        & ? &	? & NS? &	\citep{Stacey2012, Paduano2021}\\
\hline
\end{tabular}
\end{center}
\caption{Galactic globular cluster LMXBs that are persistently bright or have shown luminous outbursts (top) and faint LMXBs identified as LMXB candidates through multi-wavelength or spectral studies (bottom). Table partially based on \citep{Bahramian2014} and updated. X-ray bursts have been detected from all the bright LMXBs except for AC 211 in M15, IGR J17361-4441 in NGC 6388, CXOU J173324.6-332321  in Liller 1, and MAXI J1848-015 in GLIMPSE C01. It is important to note that there have been historical outbursts from the direction of some of these globular clusters that have been attributed to a known transient and absence of high-resolution follow up did not allow confirming such association. State: P=persistent (or active for $>$30 years), T=transient, V=variable (swinging between luminosity levels regularly).
Nature: U=ultra-compact, N=normal.  Notes: X=period from X-ray photometry, UV=period from UV photometry, XP=period from X-ray pulsations, O=period from optical photometry, B=nature of donor inferred from properties of X-ray bursts, H$\alpha$=hydrogen seen in optical counterpart spectrum. Notes represent method of measuring P$_{orb}$ or determining donor natures. $^a$- Or the alias period of 13 minutes.
}
\label{tab:gcLMXBs}
\end{table*}

\subsection{Orbital period distribution}\label{subsec:porb}
\begin{figure}
    \centering
    \includegraphics[width=10cm]{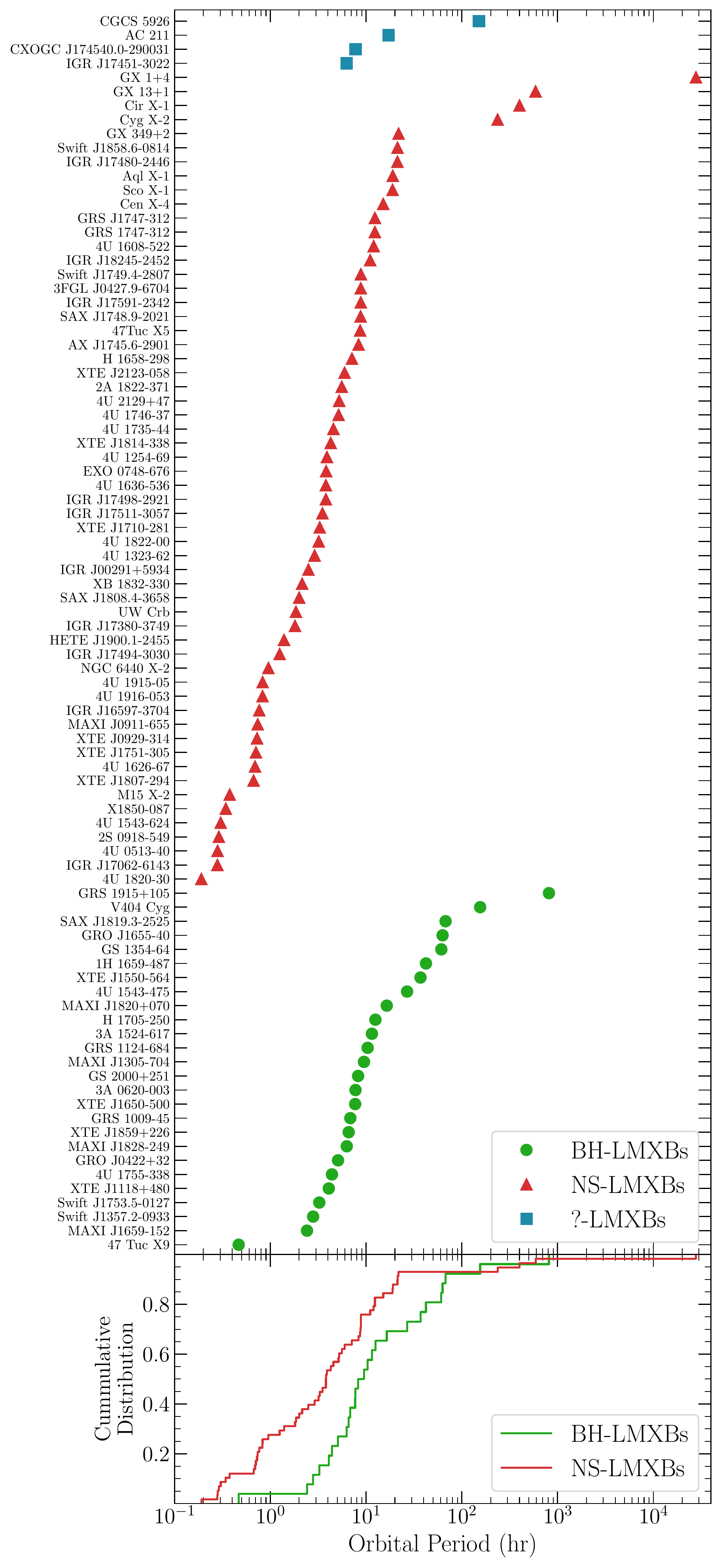}
    \caption{Top: orbital period for a subset of LMXBs. Periods for most BH-LMXBs and candidates (green circles) are from \citep{CorralSantana16} and references therein. Measurements for NS-LMXBs (red triangles) and LMXBs with unknown/uncertain compact objects (blue squares; denoted with ``?-LMXBs'') are from the literature, most of which are discussed in Sec.~\ref{sec:donor}. Note that in almost all cases uncertainties on orbital period are too small to be visible on the plotted range. Bottom: comparison between cumulative distribution of orbital periods in BH-LMXBs versus NS-LMXBs.}
    \label{fig:BH_Porb}
\end{figure}

For a limited fraction of (transient and persistent) LMXBs, the orbital period has been measured (via eclipses, radial velocity studies, pulsar timing). There are $\sim$21 confirmed BH LMXBs with orbital periods measured, which vary from a few hours to several weeks, see Figure~\ref{fig:BH_Porb}. Almost all of these systems show orbital periods of $<$1 week (with the exception of GRS~1915+105). Furthermore, there are currently no confirmed BH-LMXBs with orbital period $<4$~hr\footnote{While the ultracompact nature of 47Tuc~X9 is clear (with an orbital period of 28~minutes), BH nature of the compact object in this system is not confirmed.}. This suggests a notable paucity of BH-LMXBs with orbital period $<4$~hr \cite{arur2018} compared to theoretical predictions \cite{romani1992,portegieszwart1997,yungelson2006,repetto2015}.

While recent publications show the orbital period distribution of BHs, there is no such recent overview for NSs. Therefore, we compiled an overview of measured orbital periods in NS-LMXBs to compare them to the BH systems. The presented sample of NS-LMXBs here contains 57 systems in which the nature of the compact object was confirmed via detection of Type I X-ray bursts or pulsations, ranging in orbital periods from 11~min (4U~1820$-$30) to 3.1~yr (GX~1+4). 

Figure~\ref{fig:BH_Porb} compares the known orbital periods of the NS- and BH- LMXB populations. One aspect about the distribution that is highlighted by this figure is that we know of many NSs with orbital periods $<4$~hr, while in the BH population we know of very few of those systems \cite{Knevitt2014,arur2018}. It is possible that this is due to selection effects. For example, it could be that it is easier to find NSs with short orbital periods because the presence of a solid surface (instead of an event horizon) causes the radiative efficiency of NSs to be higher than that of BHs at low accretion rates \cite{garcia2001}.

\section{Summary and prospects for future observatories and surveys}\label{sec:summary}
In this Chapter, we reviewed observational characteristics of LMXBs linked with the properties of the compact object (NS or BH), companion star (main sequence, giant, or degenerate) in these systems, along with the binary configuration (inclination angle, Roche-lobe overflow) and accretion status. Multiple aspects of accretion and X-ray binaries from both theoretical and observational perspectives are explored in further depth in other chapters of this book. Namely, theoretical and observational details of accretion around compact objects (Chapter by Davis et al. and Chapter by Qingcui Bu), observational characteristics of black holes (Chapter by Kalemci et al. and Chapter by Garcia), evolution of X-ray binaries (Chapter by Bellono et al.), and Low magnetic field neutron stars (Chapter by Di Salvo et al.).

As indicated throughout the Chapter, the observational data is still incomplete (or biased) on many fronts and thus limits our capacity to put tight constraints on interpretations relating to population and evolution of LMXBs. However, the growing multi-wavelength focus on study of LMXBs along with the emergence of high-sensitivity wide-field astronomical surveys and facilities (such as MeerKAT in the radio, {\it Gaia} mission in the optical regime, and the {\it eROSITA} mission in the X-rays) already showcase significant advancements. These advancements include expansion in the sample of known LMXBs by dozens in the X-rays \citep[e.g., by {\it eROSITA},][]{Doroshenko2014} or identification of radio counterparts for many LMXBs (and candidates) and numerous recycled pulsars (particularly in globular cluters) by MeerKAT.

Upcoming and future missions and observatories such as (but not limited to) the next-generation VLA (ngVLA), Square-Kilometre Array (SKA) in the radio, Vera Rubin Observatory (VRO) and James Webb Space Telescope (JWST) in the infrared and optical, Imaging X-ray Polarimetry Explorer (IXPE), X-Ray Imaging and Spectroscopy Mission (XRISM), enhanced X-ray Timing and Polarimetry mission (eXTP) and Advanced Telescope for High Energy Astrophysics (ATHENA) in the X-rays promise an exciting era in exploration of Low-mass X-ray Binaries.


\section*{Acknowledgements}
The authors thank 
J.~C.~A.~Miller-Jones,
J.~M.~Corral-Santana,
T.~J.~Maccarone,  
I.~Mandel, 
J.~A.~Kennea, 
K.~C.~Dage,
and K.~Auchettl 
for helpful comments and discussions. 

AB is supported by the International Centre for Radio Astronomy Research, a joint venture between Curtin University and the University of Western Australia, funded by the state government of Western Australia and the joint venture
partners.

\clearpage


\section{Cross-References}\label{sec:xref}
Below we list a selection of other chapters in this book and section that provide more details on various aspects mentioned in this chapter.

\begin{itemize}

\item  {\it Overall accretion disk theory} \\
-- Shane Davis

\item  {\it Overall binary evolution theory} \\ 
-- Diogo Belloni \& Matthias Schreiber

\item  {\it Black holes: Accretion processes}\\
-- Qingcui Bu  

\item  {\it Black holes: Timing and spectral properties and evolution}\\
-- Emrah Kalemci, John A. Tomsick, Erin Kara  

\item  {\it Fundamental physics with black holes}\\
-- Javier Garcia

\item  {\it Low magnetic field Neutron Stars}\\
Tiziana Di Salvo, Luciano Burderi, Rosario Iaria

\end{itemize}


\bibliography{all_references}

\end{document}